\documentclass[a4paper,11pt]{article}
\pdfoutput=1 

\usepackage{jheppub} 
\usepackage[T1]{fontenc}
\usepackage{amsfonts}
\usepackage{amssymb}
\usepackage[utf8]{inputenc} 
\usepackage{graphicx} 
\usepackage{braket}
 \usepackage{fancyhdr}
\usepackage{mathrsfs}
\usepackage{setspace}
\usepackage{amsfonts}
\usepackage{amssymb}
\usepackage{amsmath}
\usepackage{bbm}
\usepackage{epsfig}
\usepackage{latexsym}
\usepackage{color}
\usepackage{nicefrac}
 \usepackage{slashed}
 \usepackage{multirow}
 \usepackage{comment}
 \usepackage{hyperref}
\usepackage{slashed}
\usepackage{array}
\usepackage{booktabs}
\usepackage{float}
\usepackage{youngtab}
\usepackage{ytableau}
\usepackage{comment}
\usepackage{mathrsfs}

\DeclareMathOperator\cn{cn}
\DeclareMathOperator\sn{sn}
\DeclareMathOperator\dn{dn}

\DeclareMathOperator\cd{cd}
\DeclareMathOperator\nd{nd}
\DeclareMathOperator\sd{sd}

\newcommand{\co}{\mathcal{O}}

\def\be{\begin{equation}}
\def\ee{\end{equation}}

\newcommand{\eps}{\epsilon}

\def\sn{{\rm sn}}
\def\cn{{\rm cn}}
\def\dn{{\rm dn}}
\def \ELE {{\mathbb{E}}}
\def \KK {{\mathbb{K}}}
\newcommand{\ba}{\begin{eqnarray}}
\newcommand{\ea}{\end{eqnarray}}

\def \T {{\cal T}}

\newcommand{\order}[1]{ \mathcal{O}\left({#1}\right) }

\title{\boldmath
Conformal Spectrum of (Super) Chiral Ising and XY Models }

\author{
Oleg Antipin$^{1}$,
Jahmall Bersini$^{2}$,
Jacob Hafjall$^{3}$,
Giulia Muco$^{3}$,
Francesco Sannino$^{3,4}$
}

\emailAdd{oantipin@irb.hr}
\emailAdd{jahmall.bersini@unibe.ch}
\emailAdd{jhafjall@qtc.sdu.dk}
\emailAdd{giulia@qtc.sdu.dk}
\emailAdd{sannino@qtc.sdu.dk}

\affiliation{
$^{1}$Rudjer Boskovic Institute, Division of Theoretical Physics,
Bijeni\v cka 54, 10000 Zagreb, Croatia
}

\affiliation{
$^{2}$Albert Einstein Center for Fundamental Physics,
Institute for Theoretical Physics, University of Bern,
Sidlerstrasse 5, CH-3012 Bern, Switzerland
}

\affiliation{
$^{3}$Quantum Theory Center ($\hslash$QTC) at IMADA \& D-IAS,
Southern Denmark University, Campusvej 55, 5230 Odense M, Denmark
}

\affiliation{
$^{4}$Dept. of Physics E. Pancini, Universit\`a di Napoli Federico II,
via Cintia, 80126 Napoli, Italy
}

\date{}
\abstract{We determine the spectrum of scaling dimensions of composite operators in the critical Gross–Neveu–Yukawa and Nambu–Jona-Lasinio–Yukawa models through a next-to-leading order semiclassical computation. Their infrared fixed points belong to the chiral Ising and chiral XY universality classes, respectively. At next-to-leading order, we resolve the Lorentz structure of the operator spectrum. For the parameter values at which supersymmetry is expected to emerge in the infrared, we obtain a closed-form expression encoding the full spectrum within the semiclassical methodology.  The resulting organization of operators into superconformal multiplets provides further evidence for emergent superconformal dynamics.
}

\arxivnumber{}

\begin{document}
\maketitle

\newpage
\section{Introduction}

\noindent
The chiral Ising and XY universality classes are relevant fermionic conformal field theories (CFTs) which respectively describe the breaking of a discrete and a continuous chiral symmetry.  One can probe these CFTs via the ultraviolet (UV) fixed points of the time-honored Gross–Neveu (GN) and Nambu–Jona-Lasinio (NJL)  models in $2 < d < 4$ dimensions~\cite{Fei:2016sgs}. Since their introduction~\cite{Gross:1974jv, Nambu:1961tp} these models stand as theoretical laboratories for studying dynamical symmetry breaking in different dimensions. The models are of further interest since, in condensed matter, they describe critical phenomena related to Dirac and Weyl semimetals~\cite{Herbut:2009qb}.   It has been conjectured~\cite{Hasenfratz:1991it,Zinn-Justin:1991yn} that the UV fixed points of the GN and NJL models fall in the same universality classes as the infrared (IR) fixed points of the corresponding Yukawa (GNY and NJLY) models with Lagrangians 
\begin{align}
    \mathcal{L}_{\text{GNY}} = \frac{1}{2}(\partial \sigma)^2 + \bar{\psi}_j i \slashed{\partial} \psi^j - y \bar{\psi}_j \sigma \psi^j - \frac{\lambda}{4} \sigma^4 \ ,
\end{align}
and
\begin{align}
    \mathcal{L}_{\text{NJLY}} = \frac{1}{2}(\partial \phi) (\partial \bar{\phi}) + \bar{\psi}_j i \slashed{\partial} \psi^j - y \bar{\psi}_j (\phi_1 + i \gamma_5 \phi_2) \psi^j - \frac{\lambda}{4} (\phi \bar{\phi})^2 \ ,
\end{align}
respectively~\cite{Fei:2016sgs}.  In this sense, these models can be viewed as UV-completions of the GN and NJL theories.  Here $\psi_j$ is a four-component Dirac fermion transforming in the fundamental of $U(N_f)$ while $\sigma$ is a $\mathbb{Z}_2$-odd real scalar and $\phi$ a complex scalar. The GNY model with  $N_f=2$ describes the critical behaviour of semimetal-insulator transitions on the graphene honeycomb lattice  
~\cite{Herbut:2006cs, Herbut:2009vu}. The case of $N_f = 1$ corresponds to spinless fermions on the same type of lattice undergoing a semimetal-insulator transition~\cite{Herbut:2009qb}. Finally,  the case with $N_f \to 0$ maps into a system where strong disorder drives relativistic semimetals through a quantum phase transition, eventually achieving a diffusive metallic phase~\cite{Louvet_2016}. The NJLY model for $N_f = 2$  describes superconducting states in graphene~\cite{Roy_2013}. A striking feature of both models is that they exhibit emergent supersymmetry (SUSY) for selected values of $N_f$~\cite{Fei:2016sgs,Thomas:2005, Zerf_2017}.  Notably, this is the case of the GNY model, which for $N_f = 1/4$ is relevant for the description of the critical point at the boundary of a $3+1$ dimensional topological superconductor~\cite{Grover:2013rc}, and the NJLY model, which for $N_f = 1/2$ describes the critical dynamics of three-dimensional topological insulators~\cite{Ponte:2012ru}. The low-lying spectrum has been determined via different methodologies, which include $\epsilon$-expansions~\cite{Fei:2016sgs, Zerf_2017, Gracey:2025aoj}, functional RG~\cite{Janssen:2014gea, Vacca:2015nta}, large $N$~\cite{Manashov:2017rrx, Gracey:2017fzu, Gracey:1993kc}, conformal bootstrap~\cite{Bashkirov:2013vya,Rong:2018okz,Iliesiu:2015qra,Atanasov:2022bpi,Iliesiu:2017nrv,Atanasov:2018kqw}, large charge~\cite{Antipin:2022naw, Bersini:2025lxs}, Lattice and Monte Carlo methods~\cite{Chandrasekharan:2013aya, Wang:2014cbw, Hands:1992be} and also the recently introduced Fuzzy Sphere methodology~\cite{Tang:2025wtj}.

In this work we use a semiclassical approach to go beyond the state-of-the-art by determining the spectrum of {\it heavy} composite operators for these CFTs at any $N_f$ in $4-\epsilon$ dimensions. This is achieved by generalizing our previous work ~\cite{Antipin:2024ekk, Antipin:2025nlq, Antipin:2025rsr}\footnote{Recent fuzzy-sphere results provide further evidence for an underlying semiclassical structure in the \(\phi^n\) operator family of the 3D Ising CFT~\cite{Zhou:2026ugg}.} to include fermions and account for Yukawa interactions around a nontrivial classical saddle emerging in the limit where the number of fields making up the composite operator is large. Notably, at the leading semiclassical order, the whole spectrum of composite operators is degenerate, and to start disentangling different Lorentz structures we are forced to include quantum fluctuations as previously noted in ~\cite{Antipin:2025nlq, Antipin:2025rsr}.

Remarkably, for the models exhibiting emergent supersymmetry, both bosonic and fermionic fluctuation operators take the form of finite-gap associated Lam\'e operators, allowing us to obtain a closed-form semiclassical result at next-to-leading order and test the expected superconformal relations.

 Furthermore, we present the spectral solution for general Darboux-Treibich-Verdier (DTV) fluctuation operators that includes the Lam\'e case and has broad applications in cosmology, string theory, particle and condensed matter physics~\cite{Greene:1997fu, Beccaria:2010ry, Togawa:2016}.  

\vskip 0.5em
 
   \noindent  \textit{Attached to this work is an ancillary Mathematica file "GNY\_NJLY\_scaling\_dimensions.nb" in which we provide explicit perturbative results up to the four-loop order for general $N_f$ and to eight loops for the values of $N_f$ for which supersymmetry emerges.}

\section{Introducing the Chiral Ising and XY Models}
The action for the GNY model in $d=4-\epsilon$ is
\begin{align}\label{eq: GNY action}
 S_{GNY} = \int d^d x \ \left[\frac{1}{2}(\partial \sigma)^2 +i \bar{\psi}_j  \slashed{\partial} \psi^j - y \bar{\psi}_j \sigma \psi^j - \frac{\lambda}{4} \sigma^4 \right] \ .
\end{align}
Here $\sigma$ is a real scalar and $\psi_j$ where $j = 1, 2, \ldots N_f$ are $N_f$ Dirac spinors transforming in the fundamental representation of $U(N_f)$. The $2$-loop beta functions are
\begin{align}
    \beta_y &= -\frac{\epsilon}{2} y  +\frac{(N+6) y^3}{32 \pi ^2} +\frac{\frac{1}{4} (-3) (4 N+3) y^5-12 \lambda  y^3+3 \lambda ^2 y}{256 \pi ^4} \ , \\
      \beta_\lambda &= -\epsilon \lambda +\frac{1}{24 \pi ^2} \left(27 \lambda ^2-3 N y^4+3 \lambda  N y^2\right) \nonumber \\ & \ \ +\frac{1}{48 \pi ^4} \left(-\frac{153 \lambda ^3}{4}+3 N y^6+\frac{21}{16} \lambda  N y^4 
    -\frac{27}{8} \lambda ^2 N y^2\right) \ ,
\end{align}
where $N = 4 N_f$ is the number of Majorana fermions when dimensionally reducing from $d=4$ to $d=3$.   Between two and four dimensions the model has an IR fixed point describing the breaking of a discrete chiral symmetry at criticality.  
The fixed point couplings are: 
\begin{align}
    \label{eq: y2fp}
    \frac{y^2_*}{(4 \pi)^2} &= \frac{\epsilon }{N+6} + \frac{\left(-N^2+(N+66) X+516 N+882\right) \epsilon ^2}{108 (N+6)^3} \ , \\
    \label{eq: lambdafp}
    \frac{\lambda_*}{(4 \pi)^2} &= \frac{1}{6} \bigg[ \frac{\left(X-N+6\right) \epsilon }{6 (N+6)}+ \frac{1}{54 (N+6)^3 X} \bigg(3 N^4+155 N^3+2745 N^2\nonumber\\
    & \ \ -X \left(3 N^3-43 N^2-1545 N-1224\right)-2538 N+7344\bigg) \epsilon^2\bigg] \ ,
\end{align}
where $X = \sqrt{N^2+132 N+36}$. In $d=4$ the global symmetry group of the model is seen to be $U(N_f) \times \mathbb{Z}_2$.

Reducing to $d=3$ the original $4$-component Dirac spinors decompose into four $2$-component Majorana fermions enhancing the global symmetry to $(O(2N_f) \times O(2N_f)) \rtimes \mathbb{Z}_2$ as also observed in~\cite{Erramilli:2022kgp, Dondi:2022zna, Zerf_2017,Osborn:1993cr}.  Throughout the paper we will generally discuss composite operators in terms of their representations of the four-dimensional Lorentz group as well as of the global $U(N_f)$. Furthermore, we will adopt a schematic notation where the composite operators in the interacting theory in $d=4-\epsilon$ are identified with their free-field theory counterpart.

The second model we investigate here is the NJLY model with action: 
\begin{align}\label{eq: NJLY action}
    S_{NJLY} = \int d^d x \ \left[\frac{1}{2} \partial^\mu \bar{\phi}\partial_\mu \phi  +i \bar{\psi}  \slashed{\partial} \psi - y \bar{\psi}_j (\phi_1 + i \gamma_5 \phi_2) \psi^j - \frac{\lambda}{4} (\phi \bar{\phi})^2 \  \right] \ ,
\end{align}
consisting of a complex scalar $\phi = \phi_1 + i \phi_2$ and $N_f$ fermions. Besides the $U(N_f)$ flavour symmetry, the model has an additional continuous $U(1)$ chiral symmetry that acts as
\begin{align}
    \phi \to e^{-2i \alpha} \phi \ , \quad \psi \to e^{i \alpha \gamma^5} \psi \ .
\end{align}
In $d=4$ the global symmetry group is therefore $U(N_f)_V \times U(1)_A$. In $d=3$ the symmetry is enhanced to $O(2 N_f) \rtimes O(2)$. The two-loop beta functions are
\begin{align}
    \beta_y &= -\frac{y}{2}  \epsilon + \frac{(N+4)  y^3}{32 \pi ^2}+ \frac{(7-3 N) y^5-16 \lambda  y^3+4 \lambda ^2 y}{256 \pi ^4} + \ldots \ ,  \\
    \beta_\lambda &= -\eps \lambda + \frac{10 \lambda ^2+N y^2 \left(\lambda -y^2\right)}{8 \pi ^2}+\frac{-120 \lambda ^3+8 N y^6+\lambda  N y^4-10 \lambda ^2 N y^2}{128 \pi ^4} + \ldots \ ,
\end{align}
At the fixed point
\begin{align}
    \frac{y_*^2}{(4\pi)^2} &=\frac{\epsilon }{N+4}+\frac{\epsilon ^2 (N (-N+Y+448)+76 Y-1096)}{100 (N+4)^3} + \ldots \ , \\
    \frac{\lambda_*}{(4\pi)^2} &=\frac{\epsilon  (-N+Y+4)}{40 (N+4)}+ \frac{3}{1000 (4 + N)^3 Y} \bigg[1280 - 26192 N + 764 N^2 \nonumber\\
    & \ + 114 N^3 + 
 3 N^4 + (320 - 3 N (-644 + (-38 + N) N)) Y\bigg]\eps^2+\ldots  \ ,
\end{align}
where we have also introduced a shorthand $Y = \sqrt{16+N(152+N)}$. In $d=3$ the model is expected to fall in the universality class of the chiral XY model.

\section{Semiclassical quantization}\label{sec: chiral ising model}

Here we will discuss the inclusion of fermions into the semiclassical approach developed in~\cite{Antipin:2024ekk, Antipin:2025nlq, Antipin:2025rsr} and pedagogically presented in~\cite{Sannino:2026ena}. For simplicity, we elucidate the semiclassical computation for the GNY model since the computations for the NJLY model are very similar. We will, however, underline the main differences when relevant.  Via the state operator correspondence, scaling dimensions of local operators in flat space become energy levels on the unit-radius cylinder $\mathbb{R} \times S^{d-1}$, where the GNY Lagrangian becomes
\begin{align}\label{eq: cylinder action}
\mathcal{L=} \ \frac{1}{2}(\partial \sigma)^2 +i \bar{\psi}_j  \slashed{\nabla} \psi^j-\frac{1}{2}\mu^2 \sigma^2 - y \bar{\psi}_j \sigma \psi^j - \frac{\lambda}{4} \sigma^4  \ .
\end{align}
Here $\mu = (d-2)/2$ is the mass arising from the conformal coupling to the curvature of $S^{d-1}$ and $\slashed{\nabla} = \gamma^a \nabla_a$ with $\nabla_a$ a covariant derivative along the vielbein $e_a = e_a^{\ \mu} \partial_\mu$ and $\gamma^a$ the flat space gamma matrices. Since the cylinder is a product manifold, we choose a time-independent orthonormal frame with $e_0^{\ \mu} = \delta_0^{\ \mu}$ so that $\nabla_0 = \partial_0$. It is convenient to introduce the resolvent $G(E)$:
\begin{align}
    G(E) = \sum_n \frac{1}{E_n-E} = i \int_0^\infty dT \ e^{i ET} \ \text{tr}e^{-i H T} \ ,
\end{align}
because it allows determining the energy spectrum from its simple poles. The semiclassical computation amounts to performing a saddle point approximation of the right-hand side, where 
\begin{align}\label{eq: saddlepoint form}
    \text{tr} \ e^{-i H T} \sim e^{i S_\text{cl}} \times \frac{\det F}{\sqrt{\det B}} \ ,
\end{align}
with $S_\text{cl}$ being the classical action of a periodic field configuration with period $\cal T$, where periodicity arises from the functional trace. The exponential yields the leading order (LO) semiclassical contribution, whereas the leading quantum correction follows from the functional determinants of fermionic $F$ and bosonic $B$ fluctuation operators subject to anti-periodic and periodic boundary conditions, respectively. Heavy neutral states on the cylinder correspond to composite operators built out of a large number $n$ of scalar fields. We consider a homogeneous classical solution of the equation of motion that we indicate with $v(t)$ scaling as $v \sim 1/{\sqrt{\lambda }} \sim 1/{\sqrt{\eps }}$. As shown in~\cite{Antipin:2025rsr, dhn} at the leading order the condition for a pole of $G(E)$ translates into the Bohr-Sommerfeld quantization condition for the classical orbits
\begin{align}\label{eq: bohr-sommerfeld-cond}
    \Omega_{d-1} \int_0^\mathcal{T}dt \  \dot{v}^2  = 2 \pi n \ ,
\end{align}
where $\T$ is the period of $v(t)$ \footnote{For a natural interpretation in terms of quantization of action variables, see~\cite{Sannino:2026ena}.}. Therefore, by considering the double scaling limit 
\begin{align}
    n\to \infty \ , \quad \eps \to 0 \ , \quad \eps n \ \text{fixed} \ ,
\end{align}
the spectrum of scaling dimensions can be organized in a semiclassical expansion in powers of $1/n$
\begin{align}
    \Delta_{n,q_\ell} = n C_0(\eps n) + C_1 \left(\eps n,\{q_\ell\} \right) + \co(1/n) \ .
\end{align}
Here $q_\ell$ are occupation numbers~\cite{Antipin:2025rsr, dhn} of the excited states on the cylinder, as will be explained below. These occupation numbers are held fixed as \(n\to\infty\), so that fluctuation excitations remain parametrically subleading to the classical background. The LO $nC_0$  stems from the classical part of eq.~\eqref{eq: saddlepoint form} and $C_1$ from the fluctuation determinants.  

\subsection{Classical result}\label{sec: classical result}
Since the classical fermion fields vanish, the relevant classical trajectory is still the one of the $\phi^4$ theory~\cite{Antipin:2024ekk} given in terms of the Jacobi elliptic cosine $\cn$ 
\begin{align} \label{eq: diclasse}
    v(t) \equiv \braket{\sigma} = \mu \sqrt{\frac{2m}{\lambda(1-2m)}} \cn(\omega t|m) \ , \quad \omega = \frac{\mu}{\sqrt{1-2m}} \ .
\end{align}
The leading semiclassical order $C_0$ corresponds to the energy $E_{cl}$ of this solution 
\begin{align} \label{eq: cl energy}
n C_0(\lambda n)  = E_{cl}  = \frac{2 \pi^2 m (1-m)}{\lambda (1-2m)^2} \ .
\end{align}
The Bohr-Sommerfeld condition eq.~\eqref{eq: bohr-sommerfeld-cond} determines the elliptic modulus $m=m(\lambda n)$ as
\begin{align}
\label{mlambdan}
\lambda n  &= \frac{8 \pi}{3(1-2m)^{3/2}} \left[(2m-1) \mathbb{E}(m) + (1-m) \mathbb{K}(m)\right] ,
\end{align}
which quantizes the classical orbit \eqref{eq: diclasse}. We proceed to include the effect of quadratic fluctuations, which introduce a splitting in the spectrum of scaling dimensions.

\subsection{Functional determinants}

We proceed by computing the functional determinants of the fluctuations around the classical solution \eqref{eq: diclasse}.
Given that the classical configuration is homogeneous, the fluctuation operators can be diagonalized in spherical harmonics labelled by an integer $\ell$. For each $\ell$, then, the bosonic operator $B(\ell)$ takes the form of a $2$-gap Lamé operator
\begin{align}
    \label{eq: scalar fluctuation}
    B(\ell) &= - \partial_u^2 + 6 m \ \sn^2(u|m) - \Lambda_B(\ell) \ , \\
    \Lambda_B(\ell) &= 6 m + (1-2m)  \left(1+\frac{\ell}{\mu}\right)^2  \ ,
        \label{Lambada}
\end{align}
where $u = \mu t/\sqrt{1-2m}$.
The scalar operator $B$ is of the Hill's form, namely a Schrödinger operator with a periodic potential. Then, according to Floquet theory, its functional determinant can be computed by finding Bloch solutions $\xi_\ell$ to the fluctuation equations of motion
\begin{align}
    B(\ell) \xi_\ell = 0 \ , \qquad
    \xi_\ell(u+4\mathbb{K}(m)) = e^{\pm i \nu_{B,\ell}} \xi_\ell(u) \ ,
\end{align}
 where $\nu_{B,\ell} \geq 0$ is the stability angle of the $\ell$'th mode. For the $2$-gap Lamé operator, the latter is known in closed form~\cite{book1, Pawellek:2008st}, which is
\be 
 \nu_{B,\ell} = 2 \pi - 4 i \KK(m) \left(\, Z(\alpha_+\,|\,m)+\, Z(\alpha_-\,|\,m) \right) \, ,
\ee
where $Z$ is the Jacobi Zeta function and $\alpha_\pm$ solve 
\begin{align}
\label{eq:Translos}
 \sn^2(\alpha_\pm \,|\,  m)=\frac{4(1+m)-\Lambda_B(\ell)}{6m}\pm \frac{1}{2m}\sqrt{\frac{4}{3}(1-m+m^2)-\frac{1}{3}\left(\Lambda_B(\ell)-2(1+m)\right)^2} \,.
\end{align}
Then, the oscillating contribution from bosons can be written in terms of a sum over the stability angles (Floquet indices)
\begin{align}
    \frac{1}{\sqrt{\det B}} \sim \sum_{\{q_B\}} e^{-i \sum_{\nu_{B,\ell}>0}(\frac{1}{2} + q_{B,\ell}) \nu_{B,\ell}} \ .
\end{align}
Here $\{q_B\}$ are bosonic occupation numbers which can take any non-negative integer value in accordance with spin statistics~\cite{dashen:1975xh}.

Since the ground state breaks only the $\mathbb{Z}_2$ part of the global symmetry\footnote{However, a $\mathbb{Z}_2$ transformation combined with a half-period time translation remains a symmetry of the ground state.}, all the fermionic flavors $\psi_j$ have the same fluctuation operator given by 
\begin{align}\label{eq: fermion operator}
    F_j = i \slashed{\nabla}  -  y v(t) \ .
\end{align}
To compute the corresponding fermionic determinant we first diagonalize the spatial part of the operator in eq.~\eqref{eq: fermion operator} by expanding the fermion fields in spherical modes. To do so, note that eigenfunctions of the Dirac operator on  $S^3$ are $2$-dimensional. Therefore we introduce two-component spinors
\begin{align}
    \psi_j = \begin{pmatrix}
        \psi^{(-)}_j\\
        \psi^{(+)}_j
    \end{pmatrix} \ .
\end{align}
They satisfy the following equation of motion
\begin{align}
    \begin{pmatrix}
        i \partial_t - y v(t) & i \slashed{\nabla}_{S^3}\\
        -i \slashed{\nabla}_{S^3} & - i \partial_t - y v(t)
    \end{pmatrix} \begin{pmatrix}
        \psi_j^{(-)}\\
        \psi_j^{(+)}
    \end{pmatrix} = 0 \ ,
\end{align}
where $\slashed{\nabla}_{S^3} = \sigma^k \nabla_k$. On $S^{d-1}$ there are two families of basis modes with the following eigenvalues and associated multiplicities~\cite{Camporesi_1996}
\begin{align}\label{eq: dirac operator eigenvalues}
    \lambda_f^{\pm} &= \pm i \left(\ell + \frac{d-1}{2} \right) \ , \\
    d_f(\ell) &=2 g_F(\ell) \,, \qquad g_F(\ell) = \frac{\Gamma(\ell+d-1)}{\Gamma(\ell+1) \Gamma(d-1)} \ .
\end{align}
The factor of $2$ arises from the dimension of the spinor representation $2^{[\frac{d-1}{2}]}$ in $d-1$ dimensions evaluated for $d=4$, while $g_F$ is the angular momentum degeneracy. 
Finally, we may combine the EOM for each of the $2$-component spinors to get a second-order equation
\begin{align}
    F_j^{(\pm)} \psi_j^{(\pm)} = 0 \ , 
\end{align}
where
\begin{align}
    \label{eq: fermionSUSYoperator}
    F_{j}^{(\pm)}(\ell) = -\partial_t^2 + \slashed{\nabla}^2_{S^3} -y^2 v^2 \pm i y \partial_t v \ .
\end{align}
Formally, this is equivalent to “squaring” the operators
\begin{align}\label{eq: squared step}
   (i \slashed{\nabla} + y v)(i \slashed{\nabla} - y v) = -\partial_t^2 + (\gamma^a \nabla_a)^2 - i y \gamma^0 (\partial_t v) - y^2 v^2= \begin{pmatrix}
        F^{(-)}_j & 0 \\
        0 & F^{(+)}_j
    \end{pmatrix} \ ,
\end{align}
where the index $a$ runs over spatial indices.
Rescaling the argument as in the bosonic case, we obtain
\begin{align}
    \label{eq: spinor flucutation}
    F_j^{(\pm)}(\ell) &=  -\partial_u^2 - 2 m \frac{y^2} {\lambda} \ \cn^2(u|m) \pm i \sqrt{2m} \frac{y}{\sqrt{\lambda}} \ \sn(u|m) \dn(u|m) - \Lambda_{F}(\ell)  \ , \\
    \Lambda_{F}(\ell) &= (1-2m)\left(1+\frac{1+2\ell}{2 \mu}\right)^2  \ .
\end{align}
The operators $F_j^{(\pm)}$ are non-Hermitian and $\mathcal{PT}$-symmetric. Moreover, since $F_{j}^{(+)}$ and $F_{j}^{(-)}$ are related by complex conjugation as well as by a half-period shift, they share the same stability angles $\nu_{F,\ell}$. This can be better appreciated by defining
\begin{align}
    W(u) = -i \sqrt{2m} \frac{y}{\sqrt{\lambda}} \cn(u|m) \ ,
\end{align}
and\begin{align}
    A_+ =  \partial_u + W \ , \quad A_- = -\partial_u + W \ .
\end{align}
Then
\begin{align}
    F^{(+)}_j(\ell) = A_+ A_- - \Lambda_F(\ell) \ , \quad F^{(-)}_j(\ell) = A_- A_+ - \Lambda_F(\ell) \ .
    \end{align}
    Defining now the operators
\begin{align}
     \mathcal{H} = \begin{pmatrix}
         A_- A_+ & 0 \\
         0 & A_+ A_-
     \end{pmatrix} \ , \quad Q_+ = \begin{pmatrix}
         0 & 0 \\
         A_+ & 0 
     \end{pmatrix} \ , \quad Q_- = \begin{pmatrix}
         0 & A_- \\
         0 & 0 
     \end{pmatrix} \ ,
\end{align}
one can immediately check that these operators generate the $1d$ SUSY algebra 
\begin{align}
    &[Q_+,\mathcal{H}] = [Q_-, \mathcal{H}] = 0 \ , \\
    & \{Q_+,Q_-\} = \mathcal{H} \ , \quad \{Q_+,Q_+\} =\{Q_-,Q_-\} =0 \ .
\end{align}
The operators on the diagonal of $\mathcal{H}$ define two superpartner Hamiltonians with superpotential $W$ and hence are iso-spectral. Since, however, they are related by simple discrete transformations, namely either complex conjugation or a half-period shift, they are self-isospectral~\cite{Dunne:1997ia,Khare_2004}. The determinant of a Hill operator subject to anti-periodic boundary conditions can be expressed in terms of its stability angles as follows~\cite{Kirsten:2003py}
\begin{align}
    \det F_j^{(\pm)}(\ell) = 4 \cos^2(\nu_{F,\ell}/2) \ .
\end{align}
Thus the full fermionic determinant is
\begin{align}
    \det F &= e^{i  N_f \sum_{\nu_{F,\ell}>0} \nu_{F,\ell}} \prod_{\nu_{F,\ell}>0} \left(1+e^{-i \nu_{F,\ell}}\right)^{2 N_f}\\
    &= e^{i N \sum_{\ell} g_F(\ell)\nu_{F,\ell}} \prod_\ell \left(1+ e^{-i \nu_{F,\ell}}\right)^{2 N g_F(\ell)} \ .
\end{align}
The product encodes the contribution from the excited fermion modes
\begin{align}
    \left(1+ e^{-i \nu_{F,\ell}}\right)^{2 N g_F(\ell)} = \sum_{q_{F,\ell}=0}^{2 N g_F(\ell)} \begin{pmatrix}
        2 N g_F(\ell) \\
        q_{F,\ell}
    \end{pmatrix} e^{-i q_{F,\ell} \nu_{F,\ell}} \ ,
\end{align}
where the fermionic occupation numbers $q_{F,\ell}$ are bounded as
\begin{align} \label{eq:bound}
    0 \leq q_{F,\ell} \leq 2 N g_F(\ell)\big|_{d=4} = 2 \times N \times \frac{(\ell+1)(\ell+2)}{2} \ ,
\end{align}
in accordance with spin-statistics. Here, the factor of two arises because the stability angles associated with $F_j^{(+)}(\ell)$ and $F_j^{(-)}(\ell)$ are equal. Unlike the bosonic one, for generic values of the couplings the fermionic fluctuation operator is not finite-gap, and its stability angles do not admit a fully closed-form expression. On the other hand, as we will see in Sec.~\ref{sec: susy section} for $N=1$, $F_j^{(\pm)}$ become $2$-gap operators with known spectrum, as a striking consequence of the emergent supersymmetry. In the non-SUSY case, we can still make progress by computing $\nu_{F,\ell}$ perturbatively in the limit of small $\eps n$ to arbitrary orders, as explained in \autoref{sec: perturbative}. In the ancillary file we provide up to four-loop order results when $N$ is arbitrary.
\subsection{Calculation of $C_1$}
Putting together all the ingredients derived above, we arrive at
\begin{align}
    \text{tr} \ e^{-i H T} \sim \exp\left[i \left(S_\text{cl} -\sum_{\nu_{B,\ell}>0} \left(\frac{1}{2} + q_{B,\ell}\right) \nu_{B,\ell}  + \sum_{\nu_{F, \ell}>0} \left(N-q_{F, \ell}\right) \nu_{F,\ell}\right)\right] \ ,
\end{align}
Following the derivation presented in~\cite{Antipin:2025rsr} we obtain our final expression for $C_1$ in terms of the bosonic and fermionic stability angles
\begin{align}
    C_1 = \delta E_{1} + \frac{1}{\T} \sum_{\nu_{B,\ell}>0}\left(\frac{1}{2} + q_{B,\ell}\right)\nu_{B,\ell} - \frac{1}{\T} \sum_{\nu_{F,\ell}>0}\left(\frac{1}{2} 2N - q_{F,\ell}\right)\nu_{F,\ell} \ ,
\end{align}
As shown in \autoref{sec: renormalize}, the renormalization procedure generates an additional $1$-loop contribution $\delta E_1$ which reads
\begin{align}
    \delta E_1 = - nC_0(\lambda n) \frac{9 \lambda  (N+6)}{8 \pi ^2 \left(-N+X+6\right)} \ .
\end{align}
The sum over stability angles is divergent and needs to be renormalized. To this end, we regularize it by subtracting the divergent terms in the $\ell\to \infty$ limit and adding them back in a $\zeta$-function regularized form. As usual, the logarithmic $1/\ell$ divergence appears as a simple pole in Dim-Reg stemming from the corresponding Zeta function $\zeta(5-d) \sim 1/\epsilon$. This $1/\epsilon$ term cancels against the one arising from the renormalization procedure outlined in \autoref{sec: renormalize}. The  bosonic contribution is the same as in the $\phi^4$ theory considered in~\cite{Antipin:2025nlq, Antipin:2025rsr} and reads
\begin{align}
    \frac{1}{2} \sum_{\ell = 1} n_{B,\ell} \nu_{B,\ell} = 2 \T + \Omega_B + \sum_{\ell = 2} \rho_{B,\ell} \ ,
\end{align}
where 
\begin{align}
n_{B,\ell}&=\frac{(2\ell+d-2)\Gamma(\ell+d-2)}{\Gamma(d-1)\Gamma(\ell+1)}\\
    \Omega_B &= -\frac{3 (m (9 m-1)-2) \mathbb{K}(m)-5 \pi  (1-2 m)^{3/2}+(24-48 m) \mathbb{E}(m)}{(1-2 m)^{3/2}} \ , \\
    \rho_{B,\ell} &= \frac{1}{2}g_{B}(\ell) \nu_{B,\ell} + \frac{1}{\ell (1-2 m)^{3/2}} \bigg[\mathbb{K}(m) \big(-2 (\ell+3) \ell^3+(9-4 \ell (\ell+1) (2 \ell (\ell+2)-1)) m^2 \nonumber\\
    &+(2 \ell (\ell+1) (2 \ell-1) (2 \ell+5)-15) m+4 \ell+6\big)+6 \left(\ell^2+\ell+1\right) (2 m-1) \mathbb{E}(m)\nonumber\\
    &+\pi  \ell (\ell+1)^2 (1-2 m)^{3/2}\bigg] \ , \\
    \label{eq: boson pole}
     B_{\text{pole}} &= -\frac{3 ((1-m) (2-3 m) \mathbb{K}(m)-(2-4 m) \mathbb{E}(m))}{(1-2 m)^{3/2} \epsilon } \ .
\end{align}
Additionally, for each fermion flavor, we have
\begin{align}
    \frac{1}{2} \sum_{\ell=0} n_{F,\ell} \nu_{F,\ell} = \frac{\nu_{F,0}}{2} + \Omega_F + \sum_{\ell = 1}^\infty \rho_{F,\ell} \ ,
\end{align}
where
\begin{align}
n_{F,\ell}&=g_{F,\ell}=\frac{\Gamma(\ell+d-1)}{\Gamma(d-1)\Gamma(\ell+1)} \ , \\
    \Omega_F & = -\frac{\pi }{2}+\frac{1}{16} \pi  \left(18-7 \frac{y^2}{\lambda}\right) m-\frac{1}{128} 9 \left(\pi  \left(7 \frac{y^2}{\lambda}-13\right)\right) m^2 + \ldots \ , \\
    \label{eq: Fpoleperturb}
    F_{pole} &= \frac{3 \pi  m^2 N \left(\lambda  y^2-y^4\right)}{16 \lambda ^2 \epsilon } + \ldots \,,
    \end{align}
    and
    \begin{align}
    \rho_{F,\ell} & = \frac{1}{64 \ell (1 + \ell)^2 (2 + \ell)^2} \bigg[-3 (1 + \ell) (2 + \ell) (4 + 3 \ell) \pi \frac{y^2}{\lambda} + (24 + 
    \ell (51 + \ell (37 + 9 \ell))) \pi \frac{y^4}{\lambda^2}\bigg] m^2 + \ldots \,. 
\end{align}
\noindent
Combining the above results, we arrive at our final expression for $C_1$
\begin{align}\label{eq: chiral ising C1}
    C_1 &= \delta E_1 + 2 + \frac{\Omega_B - 2 N \Omega_F}{\T} - \frac{N\nu_{F,0}}{\mathcal{T}}  + \frac{1}{\T}\bigg[\sum_{\ell = 2} \rho_{B,\ell} - 2 N \sum_{\ell=1}\rho_{F,\ell}\bigg]\nonumber\\
    &+ \frac{1}{\T} \sum_\ell\left( q_{B,\ell} \nu_{B,\ell} + q_{F,\ell} \nu_{F,\ell} \right) \ .
\end{align}
Note that the constant term cancels when all the contributions are combined. Combining equations \eqref{eq: boson pole}, \eqref{eq: Fpoleperturb} and \eqref{eq: action pole} we see that the $\eps$-pole cancels. One can then systematically expand the above expression in the small $\lambda n$ limit to arbitrarily high orders. The corresponding composite operators are labelled by specific sets of occupation numbers $q_{B,\ell}$ and $q_{F,\ell}$ as we proceed to show in the next section.

\subsubsection{The NJLY story}
For the NJLY model we expand around a classical saddle with zero $U(1)$ charge given by
\begin{align}
    \phi_{cl}(t) = e^{i \chi_0} v(t) \ ,
\end{align}
where $\chi_0$ is a constant real phase and $v(t)$ a real homogeneous time-periodic classical background. The EOM for $v(t)$ is again that of the quartic oscillator and the leading semiclassical result is therefore given by $n C_0$ as already calculated in Sec.~\ref{sec: classical result}. We can always rotate away the phase $\chi_0$ such that fluctuations around the classical saddle are parameterized as
\begin{align}
    \phi = v(t) + (\eta_\parallel(x) +i \eta_\perp(x)) \ , \quad \psi_j = 0 + \psi_j(x) \ . 
\end{align}
The fluctuation operator for $\eta_\perp$ is a one-gap Lamé operator identical to that found in the $\phi^4$ $O(N)$ analysis of~\cite{Antipin:2025rsr} while those of $\eta_\parallel$ and the fermionic ones are identical to those for the GNY model if written in terms of the couplings. At the fixed point, the operators necessarily differ. We can therefore write down the fluctuation contribution by including the effect of the $1$-gap Lamé fluctuations $\eta_\perp$ in eq.~\eqref{eq: chiral ising C1} to obtain 
\begin{align}\label{eq: NJLY C1}
    C_1 &= \delta E_1 + 2 + \frac{\Omega_{\parallel} + \Omega_{\perp} - 2 N \Omega_F}{\T} - \frac{N\nu_{F,0}}{\mathcal{T}}  + \frac{1}{\T}\bigg[\sum_{\ell = 2} \rho_{\parallel,\ell} + \sum_{\ell = 1}\rho_{\perp, \ell} - 2 N \sum_{\ell=1}\rho_{F,\ell}\bigg]\nonumber\\
    &+ \frac{1}{\T} \sum_\ell\left( q_{\parallel,\ell} \nu_{\parallel,\ell} + q_{\perp,\ell} \nu_{\perp, \ell} + q_{F,\ell} \nu_{F,\ell} \right) \ ,
\end{align}
where the new contributions stemming from $\eta_\perp$ and renormalisation are

\begin{align}
    \delta E_1 & = - n C_0(\lambda n) \frac{5 \lambda  (N+4)}{4 \pi ^2 \left(Y-N+4\right)} \ , \\
    \Omega_{\perp} &= \pi + \frac{2 (m \KK(m)-\ELE(m))}{\sqrt{1-2 m}} \ , \\
    \rho_{\perp,\ell} &= \frac{1}{2}n_{B,\ell} \nu_{\perp,\ell}|_{d=4} +\frac{1}{3\ell(1-2m)^{3/2}} \bigg[3 \pi  \ell (\ell+1)^2 (1-2 m)^{3/2}\nonumber\\
    &\ \ +2 (3 \ell (\ell+1)+1) (2 m-1) \ELE(m)\nonumber\\
    &\ \ - \bigg(6 (\ell+1) (\ell+2) \ell^2+(5-6 \ell (\ell+1) (4 \ell (\ell+2)+1)) m-2\nonumber\\
    &\ \ +3 (4 \ell (\ell+1) (2 \ell (\ell+2)+1)-1) m^2\bigg)\mathbb{K}(m)\bigg] \ ,  \\
    B_{\text{pole},\perp} &= -\frac{(m-1) (3 m-2) \mathbb{K}(m)+(4 m-2) \mathbb{E}(m)}{3 (1-2 m)^{3/2} \epsilon } \ .
\end{align}
Here $ \nu_{\perp,\ell}$ denotes the stability angle of the $1$-gap Lamé operator obtained in the $\ell$'th angular momentum channel. In the upcoming section we present explicit perturbative results for the scaling dimensions of various composite operators.

\subsection{Explicit results: GNY model}
\label{sec:explicit_results}
We here provide explicit results for the operator spectrum of the theory while outlining the correspondence between composite operators and occupation numbers. By setting $q_B = q_F = 0$ we obtain the ground state energy on the cylinder which, at small $\lambda n$, can be identified with the $\sigma^n$ operator

\begin{align}\label{eq: semiclassical groundstate}
    \Delta_{\sigma^n} &= n + \frac{1}{12(N+6)}\bigg[\textcolor{red}{n (6-N+X)}+\textcolor{blue}{(N-42-X)}\bigg] \epsilon n - \frac{1}{648 (N+6)^2} \bigg[\textcolor{red}{17 n (N (N-X+60)}\nonumber\\
    &\textcolor{red}{+6 (X+6))} + \textcolor{blue}{\frac{2}{(6+N)X} \bigg(N (N (17 N (2 N-2 X+315)-1491 X+11637)}\nonumber\\
    &\textcolor{blue}{-9 (1327 X+8610))-7776 (X+6)\bigg) }+ \ldots \bigg] (\epsilon n)^2 + \ldots ,
\end{align}
where the colors red and blue denote $C_0$ and $C_1$ contributions respectively. Additional results up to four-loop order can be found in the ancillary file. Since at $1$-loop $\sigma^n$ does not mix with other operators, its $\mathcal{O}(\epsilon)$ dimension can also be determined by fitting the results for $n=1$ and $n=2$~\cite{Fei:2016sgs} to the simple polynomial ansatz $ \Delta_{\sigma^n} =  n + \left(a_1 n^2 + a_2 n \right) \epsilon$ which leads to $a_1 = (6-N+X)/12(N+6)$ and $a_2 = (N-42-X)/12(N+6)$, in agreement with our calculation. However, as a consequence of operator mixing, from the two-loop order on the dependence on $n$ becomes non-polynomial. Accordingly, the expression in \eqref{eq: semiclassical groundstate} captures the leading two terms of the large $n$ expansion of the function of $n$ appearing at any loop order. Indeed, we can further compare our results to $1/N$ calculations. The scaling dimension of $\sigma^n$ is known up to order\footnote{Notice that in the $1/N$ expansion, operator mixing for the $\sigma^n$ family starts only at order $\order{1/N^3}$~\cite{Manashov:2017rrx} and so, to this order, the $n$-dependence is polynomial.} $1/N^2$ in arbitrary $2<d<4$~\cite{Manashov:2017rrx}. Expanding their expression in $d=4-\epsilon$ we obtain
\begin{align}\label{eq: large N known}
    \Delta_{\sigma^n} &= n + \bigg[\frac{\textcolor{red}{6n^2}-\textcolor{blue}{9n}}{N}-\frac{\textcolor{red}{216 n^2} - \textcolor{blue}{234 n}}{N^2} + \ldots \bigg] \epsilon \nonumber\\
    &+ \bigg[\frac{\frac{27 n}{4}-\textcolor{blue}{5 n^2}}{N}+\frac{\textcolor{red}{-68 n^3} + \textcolor{blue}{498 n^2}-\frac{1771 n}{4}}{N^2} + \ldots \bigg] \epsilon^2 + \bigg[\frac{n (19-8 n)}{16 N} \nonumber\\
    &+\frac{\textcolor{blue}{n^3 (320 \zeta (3)+1568)}+n^2 (672 \zeta (3)-4824)+n (2975-1184 \zeta (3))}{16 N^2} + \ldots \bigg]\epsilon^3 + \ldots \ ,
\end{align}
in agreement with our results. Here, we have also color-coded terms that overlap with the semiclassical expansion. Terms that are not color coded are NNLO and higher semiclassical terms. At order $\eps^3$ the leading semiclassical order contributes from $\co(1/N^3)$, which can be seen by expanding the four-loop semiclassical result found in the ancillary file in $1/N$. The NLO, however, starts from $\co(1/N^2)$ and indeed matches the $n^3$ coefficient in the known result eq.~\eqref{eq: large N known}.

The scaling dimensions corresponding to the excited states are obtained by considering nonzero occupation numbers. This amounts to adding to the ground state expression combinations of the stability angles, whose perturbative expressions for the $\ell=0$ and $\ell \neq 0$ modes read
\begin{equation}
    \label{eq: fermion zero mode}
    \begin{split}
    \frac{\nu_{F,0}}{\T} &= \frac{1}{2}+\frac{N-X+12}{6 N+36} (\epsilon n) + \frac{12 + N - X}{648 (N+6)^2 (N-X+6)} \bigg[215X\\
    &\ \ -690 + N (3151 + 51 N - 51 X) \bigg] (\epsilon n)^2 + \ldots  , \\
    \end{split}
\end{equation}
\begin{align}
    \label{eq:ferm_excit}
    \frac{\nu_{F,\ell}}{\T} &= \frac{1}{2}+\ell + \frac{6 - N + X}{18 (6 + N)} \bigg[\frac{36 (2 \ell+3)}{\bigg(\ell^2+3 \ell+2\bigg) (-N+X+6)}-3\bigg] (\eps n)  \nonumber\\
    &\ \ + \frac{(-N+X+6)^2}{1296 (6 + N)^2} \bigg[\frac{216 (2 \ell+3) (2 \ell (\ell+3)+3)}{(\ell+1)^2 (\ell+2)^2 (N-X-6)} \nonumber\\
    &\ \ -\frac{2592 (2 \ell+3) (3 \ell (\ell+3)+7)}{(\ell+1)^3 (\ell+2)^3 (-N+X+6)^2}+51\bigg] (\eps n)^2 + \ldots  \ , \quad \ell \geq 1 \ ,
\end{align}
and
\begin{align}
    \frac{\nu_{B,\ell}}{\T} &= \ell -\frac{(\ell-1) (-N+X+6)}{6 ((\ell+1) (N+6))} (\epsilon n) \nonumber\\
    &\ \ + \frac{(\ell-1) (\ell (\ell (\ell (17 \ell+78)+135)+98)+12)  (-N+X+6)^2}{432 \ell (\ell+1)^3 (\ell+2) (N+6)^2} (\epsilon n)^2 + \ldots , \ \ell \geq 2 \ .
\end{align}
Note that $\nu_{B,1} =\T$  and $\nu_{B,0} = 0$ are descendant and zero modes, respectively, with the latter arising from the spontaneous breaking of time translation invariance on the cylinder. In~\cite{Antipin:2025nlq,Antipin:2025rsr} it was shown that exciting a bosonic $\ell$-mode corresponds to adding 
$\ell$ partial derivatives in the $\left(\ell/2,\ell/2\right)$ representation of the Lorentz group.  For instance, exciting a \textit{single} bosonic $\ell=s>1$ mode by setting $q_{B,\ell} = \delta_{\ell s}$ gives the family of spin-$s$ (traceless-symmetric) primary operators of the schematic form $\partial^s \sigma^n$ with dimension
\begin{align}
\Delta_{\partial^s \sigma^n} &=  n + s + \frac{1}{12 (N+6) (s+1)} \bigg[ n  (6 - N + X)  (1 + s) + N (-1 + 3 s) \nonumber\\
&-30 + X - 3  s (18 + X) + \ldots \bigg] \epsilon n + \frac{1}{1296 (N+6)^3} \bigg[-34 n (N+6) \big(N (N-X+60)\nonumber\\
&+6 (X+6)\big) + \frac{1}{X} \bigg(4  (7776  (6 + X) + 
   N (9  (8610 + 1327  X) + 
      N (17 N (-315 - 2 N + 2  X) \nonumber\\
      &+ 3  (-3879 + 497  X))))\bigg) + \frac{1}{s (s+1)^3 (s+2)} \bigg(3 (6 + N)  (6 - N + X)^2  (-1 + s)  \big(12 \nonumber\\&+ 
   s (98 + s (135 + s (78 + 17 s)))\big)\bigg) +  \ldots \bigg] (\epsilon n)^2 +  \ldots 
\end{align}
On the other hand, since $\nu_{B,\ell=1} =\T $, exciting an $\ell=1$ mode increases the dimension by one unit, producing conformal descendants. Analogously, fermionic $\ell$-modes are associated with the $\Big(\frac{\ell+1}{2}, \frac{\ell}{2} \Big) \ \oplus \  \Big(\frac{\ell}{2}, \frac{\ell+1}{2} \Big)$ Lorentz representations. Indeed, the dimension of these representations is exactly $2(\ell+1)(\ell+2)$ and the upper bound on the fermionic occupation number in eq.~\eqref{eq:bound} corresponds to the maximum number of fermionic fields that can be used to build a composite operator in the free field theory limit, schematically of the form
\begin{align}
    \psi^{q_{F,0}} (\partial \psi)^{q_{F,1}} (\partial^2 \psi)^{q_{F,2}} \ldots (\partial^\ell \psi)^{q_{F,\ell}} \ .
\end{align}
In this expression it is understood that out of the four possible derivatives, only a combination of three of them should appear due to the Dirac equation. Furthermore, the powers indicate only the number of fermionic degrees of freedom appearing at each derivative level, bounded by the dimension of the representation $\Big(\frac{\ell+1}{2}, \frac{\ell}{2}\Big) \ \oplus \ \Big(\frac{\ell}{2} , \frac{\ell+1}{2}\Big)$. The true primary operators are linear combinations of terms with all possible spinor contractions and gamma matrix products. Intuitively, exciting a fermionic $\ell$-mode turns scalar operators into fermionic ones. As evident from eq.~\eqref{eq:ferm_excit}, exciting a fermionic $\ell$-mode adds $\ell+1/2$ to the classical operator dimension.

It is therefore clear that the whole operator spectrum can be accessed by exciting multiple fermionic and bosonic $\ell$-modes, with the resulting operators identified by the irreducible representations appearing in the decomposition of the tensor product of the corresponding Lorentz representations. It is important to stress that at the leading semiclassical order the full operator spectrum is degenerate. While some degeneracies are lifted at the next-to-leading order, the spectrum remains highly degenerate as operators appearing in the decomposition of the same product representation still have the same dimension. Finally, at this order operators in different $U(N_f)$ representations remain fully degenerate. We proceed by providing some additional concrete examples

\begin{itemize}
    \item $q_{F,\ell}=\delta_{\ell,0}$: From eq.~\eqref{eq: fermion zero mode} we see that exciting the $\ell = 0$ fermionic mode once adds $1/2$ to the canonical dimension of $\sigma^n$ and produces operators in the $\Big(\frac{1}{2}, 0 \Big) \oplus \Big(0, \frac{1}{2} \Big)$ and $\Big(0, \frac{1}{2} \Big) \oplus \Big(\frac{1}{2}, 0 \Big)$, which can respectively be identified with $\sigma^{n-1} \psi $ and $ \sigma^{n-1} \bar{\psi} $ with equal dimension
    \begin{align}
    \Delta_{\sigma^{n-1} \psi} &= n+\frac{1}{2} + \frac{(n-3)  (-N+X+6)\epsilon n}{12 (N+6)} + \frac{1}{648 (N+6)^3} \bigg[17 n (6 + N) (N (X-60-N) \nonumber\\
    &- 6 (6 + X)) + \frac{2}{X (N-X+6)} \bigg(-540 (-654 + 89 X) \nonumber\\
    &+ 
 N (36 (717 + 713 X) + 
    N (-2884518 + 68655 X \nonumber\\
    &+ 
       N (-539151 + 12496 X + N (-20350 - 119 N + 119 X))))\bigg)\bigg](\eps n)^2 + \ldots \ .
\end{align}

\item $q_{F,\ell} = 2 \delta_{\ell,0}$: we consider the product of two Dirac representations, which is 
\begin{align}\label{eq: Dirac square representations}
    (0,0) \ \oplus \ (0,0) \ \oplus \  \Big(\frac{1}{2}, \frac{1}{2} \Big) \ \oplus \   \Big(\frac{1}{2} , \frac{1}{2} \Big) \ \oplus \  \left\{(1,0) \ \oplus \  (0,1)\right\} \ .
\end{align}
The schematic form of the corresponding operators is determined by noting that, heuristically, since the canonical dimension increases by one unit, here two scalars are turned into fermions and no derivatives are added. For instance, in the $N_f=1$ case this construction leads to the following operators
\begin{align}\label{eq: fermion bilinears}
   \sigma^{n-2} \psi^a \Gamma^{A \ b}_{\ a} \psi_b \ , \quad \sigma^{n-2}  \bar{\psi}^a \Gamma^{A \ b}_{\ a} \bar{\psi}_b \ , \quad \sigma^{n-2} \bar{\psi}^a \Gamma^{A \ b}_{\ a} \psi_b \ .
\end{align}
and $\psi^a$ is obtained by raising the spinor index with the charge-conjugation matrix~\cite{Dreiner:2008tw}. The first two operators carry $U(1)$ charge $\pm 2$ and $\Gamma^A$ denotes collectively either $\mathbb{I}, \ \gamma^5, \ \gamma^\mu \gamma^5$. Instead, for the neutral operator $\sigma^{n-2} \bar{\psi}^a \Gamma^{A \ b}_{\ a} \psi_b $, $\Gamma^A$ can be any of the $16$ independents $\gamma$-matrices bilinears  $\mathbb{I}, \ \gamma^5, \ \gamma^\mu , \ \gamma^\mu \gamma^5 , \ \sigma^{\mu \nu}=\frac{i}{2}[\gamma^\mu, \gamma^\nu]$. We stress that all operators in eq.~\eqref{eq: fermion bilinears} share the same scaling dimension at this semiclassical order, which reads
\begin{align}
    \Delta_{\sigma^{n-2}\psi^2} &=  n + 1 +\frac{(n (-N+X+6)+5 N-5 X+6) \eps n}{12 (N+6)} \nonumber\\
    &\ \ + \frac{1}{648 (N+6)^3}\bigg[-17 n (6 + N) (N (60 + N - X) + 6 (6 + X)) \nonumber\\
    &\ \ + \frac{4}{X (6 + N - X)} \bigg(-540 (-654 + 89 X) + 
 N (252 (1977 + 98 X) \nonumber\\
 &\ \ + 
    N (-2069388 + 52392 X + 
       N (-412278 + 8971 X \nonumber\\
       &\ \ + N (-14581 - 85 N + 85 X))))\bigg)\bigg](\eps n)^2 + \ldots \ .
\end{align}

\item $q_{F,\ell}=\delta_{\ell,s\ge1}$: Exciting a single fermionic $\ell=s$ mode we find operators of the schematic form
\begin{align}
    \sigma^{n-1} \partial^s \psi \quad \text{and} \quad \sigma^{n-1} \partial^s \bar{\psi} \ ,
\end{align}
which transform in the representations $\Big(\frac{s+1}{2}, \frac{s}{2} \Big) \ \oplus \  \Big(\frac{s}{2}, \frac{s+1}{2} \Big)$ and $\Big(\frac{s}{2}, \frac{s+1}{2} \Big) \ \oplus \  \Big(\frac{s+1}{2}, \frac{s}{2} \Big)$ and have scaling dimension given by
\begin{align}
    \Delta_{\sigma^{n-1}\partial^s \psi} &= n + \frac{1}{2}+s + \frac{1}{12 (6 + N) (1 + s) (2 + s)} \bigg[n (2 + 3 s + s^2) (6 - N + X) \nonumber\\
    &- 
 3 (-N (2 + 3 s + s^2) + 2 (6 + X) + s^2 (18 + X) + s (38 + 3 X))\bigg]\eps n \nonumber\\
 &+ \frac{1}{1296 (N+6)^3} \bigg[-34 n (6 + N) (N (60 + N - X) + 6 (6 + X)) \nonumber\\&
 -\frac{3 (6 + N)}{(1 + s)^3 (2 + s)^3} \bigg(25 (1 + s)^3 (2 + s)^3 (6 - N + X)^2 \nonumber\\
 &+ 
 42 (1 + s)^2 (2 + s)^2 (6 - N + X) (36 + 
    24 s - (1 + s) (2 + s) (6 - N + X)) \nonumber\\
    &+ 
 72 (3 + 2 s) (84 + 
    36 s (3 + s) - (1 + s) (2 + s) (11 + 5 s (3 + s)) (6 - N + X))\bigg) \nonumber\\
 &+ \frac{4}{X} \bigg(7776 (6 + X) + 
 N (9 (8610 + 1327 X) + N (17 N (-315 - 2 N + 2 X) \nonumber\\
 &+ 3 (-3879 + 497 X)))\bigg)\bigg](\eps n)^2 + \ldots \ .
\end{align}
Having emphasized the general construction of the excited spectrum, we move on to state explicit results for the NJLY model as well.

\end{itemize}


\subsection{Explicit results: NJLY model}
\label{sec:explicit_results_NJLY}
The semiclassical ground state is
\begin{align}\label{eq: NJLY ground state}
    \Delta_{(\bar{\phi}\phi)^{n/2}} &= n + \frac{1}{40(N+4)} \bigg[3 n (4 - N + Y)+2 (N-44- Y)\bigg] \eps n \nonumber\\
    &\ \ - \frac{1}{800(4+N)^2} \bigg[17 (N (72 + N - Y) + 4 (4 + Y)) n + \nonumber\\
    &\ \ - \frac{8}{15 (4 + N) Y (-28 - 3 N + 3 Y)} \bigg(N (N (21708688 + N (3596772 + N (137278 \nonumber\\
    &\ \ + 667 N - 667 Y) - 86586 Y) - 
       437196 Y) \nonumber\\
       &\ \ + 1024 (12088 - 1813 Y)) - 111360 (4 + Y)\bigg)\bigg] (\eps n)^2 + \ldots \ .
\end{align}
As was the case for the GNY model, the absence of mixing at one-loop makes the semiclassical result exact at that order. Naturally, we expect the semiclassical ground states to correspond to the family of $O(2)$ singlet operators $(\bar{\phi} \phi)^{n/2}$ for even $n$. Indeed fitting known one-loop operator dimensions~\cite{Fei:2016sgs} of the operators $\bar{\phi}\phi$ and $(\bar{\phi}\phi)^2$ to the ansatz $\Delta_n = n + (an^2 + bn)\eps$ we obtain $a = 3 (4 - N + Y)/40 (4 + N)$ and $b=(N-44-Y)/20(4+N)$ matching eq.~\eqref{eq: NJLY ground state}. At higher orders mixing occurs, and again the semiclassical expansion determines the leading and sub-leading $n$ at all perturbative orders. The excited spectrum is obtained by exciting various $\ell$-modes of the stability angles below 
\begin{align} \label{NJLYANGLE}
    \frac{\nu_{F,0}}{\T} &= \frac{1}{2} + \frac{3 (N-Y+16)}{20 (N+4)} (\eps n) + \frac{1}{1200 (N+4)^2 (3 N-3 Y+28)} \bigg[N (N (459 N-459 Y+55072)\nonumber\\
    &\ \ -20188 Y+195812)+4 (4139 Y-38444)\bigg] (\eps n)^2 + \ldots \ , \\
    \frac{\nu_{F,\ell}}{\T} &= \frac{1}{2} + \ell + \frac{1}{20 (2 + 3 \ell + \ell^2) (4 + N)} \bigg[3 \ell^2 N-3 \ell^2 Y-12 \ell^2+9 \ell N-9 \ell Y+44 \ell\nonumber\\
    &\ \ +6 N-6 Y+96\bigg](\eps n) + \frac{1}{800 \left(\ell^2+3 \ell+2\right)^3 (N+4)^2} \bigg[51 \ell^6 (N (N-Y+72)\nonumber\\
    &\ \ +4 (Y+4))+3 \ell^5 (N (153 N-153 Y+11176)+452 (Y+4))\nonumber\\
    &\ \ +9 \ell^4 (N (187 N-187 Y+13864)+348 (Y+4)) \nonumber\\
    &\ \ +3 \ell^3 (N (1071 N-1071 Y+80552)+844 Y+176)+18 \ell^2 (N (187 N-187 Y+14244)\nonumber\\
    &\ \ -32 (Y+79))+4 \ell (9 N (51 N-51 Y+3922)-2 (207 Y+9028))+408 N (N-Y)\nonumber\\
    &\ \ +48 (657 N-11 Y-744)\bigg] (\eps n)^2 + \ldots \ ,
\end{align}
and
\begin{align}
    \frac{\nu_{B,\parallel}}{\T} &= \ell -\frac{3 (\ell-1) (-N+Y+4)}{20 (\ell+1) (N+4)} (\eps n)\nonumber\\
    &\ \ + \frac{3 \left(17 \ell^5+61 \ell^4+57 \ell^3-37 \ell^2-86 \ell-12\right) (-N+Y+4)^2}{1600 \ell (\ell+1)^3 (\ell+2) (N+4)^2} (\eps n)^2 + \ldots \ , \\
    \frac{\nu_{B,\perp}}{\T} &= \ell -\frac{(3 \ell+1) (-N+Y+4)}{20 (\ell+1) (N+4)} (\eps n) \nonumber\\
    &\ \ + \frac{(3 \ell (\ell+2) (\ell (\ell (17 \ell+43)+35)+5)-4) (-N+Y+4)^2}{1600 \ell (\ell+1)^3 (\ell+2) (N+4)^2} (\eps n)^2 + \ldots \ .
\end{align}
Again $\nu_{B,\parallel,0} = \nu_{B,\perp,0} = 0$ and $\nu_{B,\parallel, 1} = \T$. Some examples are
\begin{itemize}
    \item $q_{F,\ell} = \delta_{\ell,0}$:
\begin{align}
        \Delta_{(\bar{\phi} \phi)^{\frac{n-2}{2}} \phi \psi} &= \Delta_{(\bar{\phi} \phi)^{\frac{n-2}{2}} \bar{\phi} \psi} = n + \frac{1}{2} + \bigg[\frac{3 n (-N+Y+4)}{40 (N+4)}+\frac{N-Y+1}{5 (N+4)}\bigg]  (\eps n) \nonumber\\
        &+ \frac{1}{3000 (N+4)^3 Y (3 N-3 Y+28) (N-Y-4)} \bigg[1720320 (4 + Y) \nonumber\\
        &+ 
 N (127474432 - 
    N (560948160 + 
       N (306920560 + 
          N (63626900 \nonumber\\
          &+ N (1186740 + 4963 N)))) + (-20302912 + 
       N (66155392 + N (16394388 \nonumber\\
       &+ N (809552 + 4963 N)))) Y) + 
 255 n (4 + N) (512 (4 + Y) \nonumber\\
 &+ 
    N (64 (260 + 27 Y) + 
       N (N (33920 + N (680 + 3 N - 3 Y) \nonumber\\
       &- 452 Y) - 144 (160 + 57 Y))))\bigg] (\eps n)^2 + \ldots \ .
    \end{align}
    \item $q_{F,\ell} = \delta_{\ell,s\ge1}$:
    \begin{align}
        \Delta_{(\bar{\phi} \phi)^{\frac{n-2}{2}} \phi \partial^s\psi} &= \Delta_{(\bar{\phi} \phi)^{\frac{n-2}{2}} \bar{\phi} \partial^s\psi} = n + \frac{1}{2} + s + \bigg[\frac{3 n (-N+Y+4)}{40 (N+4)}\nonumber\\
        &+\frac{N s^2+3 N s+2 N-s^2 Y-14 s^2-3 s Y-22 s-2 Y+2}{5 (N+4) \left(s^2+3 s+2\right)}\bigg] (\eps n) \nonumber\\
        &+ \frac{1}{12000 (N+4)^3} \bigg[-\frac{510 (N+4) n}{28 + 3 N - 3 Y} \bigg(N (144 + N (344 + 3 N - 3 Y) \nonumber\\
        &- 116 Y) + 32 (4 + Y)\bigg) + \frac{80 (4 + N)}{28 + 3 N - 3 Y} \bigg(N (1048 + N (8408 \nonumber\\
        &+ 91 N - 91 Y) - 1492 Y) + 264 (4 + Y)\bigg) \nonumber\\
        &+ \frac{108}{Y} \bigg(320 (4 + Y) + 
 N (N (764 + 3 N (38 + N - Y) + 114 Y) \nonumber\\
 &+ 4 (-6548 + 483 Y))\bigg) + \frac{15 (N+4)}{\left(s^2+3 s+2\right)^3} \bigg(51 N^2 (1 + s)^3 (2 + s)^3 \nonumber\\
 &+ 
 24 N (1 + s) (2 + s) (657 + s (1956 + s (2079 + s (938 + 153 s)))) \nonumber\\
 &- 
 51 N (1 + s)^3 (2 + s)^3 Y - 48 (744 + 11 Y) + 
 4 s (-2 (9028 + 207 Y) \nonumber\\
 &+ 
    3 s (-48 (79 + Y) + 
       s (44 + 211 Y + s (261 \nonumber\\
       &+ s (113 + 17 s)) (4 + Y))))\bigg)\bigg] (\eps n)^2 + \ldots \,.
    \end{align}
\end{itemize}

\section{Emergent supersymmetry}\label{sec: susy section}

In this section, we consider the $N=1$ limit of the GNY model as well as the $N=2$ limit of the NJLY model, which are believed to feature emergent supersymmetry in three dimensions. Notably, the supersymmetric relation between quartic and Yukawa coupling turns the fermionic fluctuation operator into a finite-gap operator of the associated Lamé type whose spectrum can be exactly determined. This will allow us to provide nontrivial tests of supersymmetric relations between operator dimensions supporting the emergence of supersymmetry in the IR.

Concretely, the three-dimensional $N=1$ GNY model is expected to fall in the same universality class as the $\mathcal{N}=1$ super-Ising model~\cite{Thomas:2005} which occurs at the boundary of topological superconductors~\cite{Grover:2013rc}. In $d=3$, its action can be written in terms of the scalar superfield $\Sigma$
\begin{equation}
    \Sigma(x,\theta)=\sigma+\theta^{\alpha}\psi_{\alpha}-\theta^2 F \ ,
\end{equation}
leading to
\begin{equation}
    \mathcal{S}=- \int d^3 x d^2\theta \Bigg[\frac{1}{2}(D_\alpha\Sigma)^2+\frac{g_1}{3!}\Sigma^3\Bigg]\ .
\end{equation}
Here $\theta^{\alpha} \ (\alpha=1,2)$ denotes the Grassmann superspace coordinate, $\psi^{\alpha}$ is a Majorana fermion and $F$  is a scalar auxiliary field. The superspace covariant derivative is~\cite{Gates:1983nr} 
\be 
D_\alpha\equiv \partial_\alpha+\frac{i}{\sqrt{2}}\theta^{\beta}(\gamma^\mu)_{\alpha\beta}\partial_\mu \ .
\ee
Integrating over the superspace coordinates and integrating out $F$ via its equations of motion $F\sim\sigma^2$, the action reduces to
\begin{equation}
    \mathcal{S}=\int d^3x \Bigg(-\frac{1}{2}\sigma\Box\sigma+\frac{i}{2}\psi \slashed{\partial} \psi -\frac{g_1}{2}\sigma \psi\psi -\frac{g_1^2}{8}\sigma^4\Bigg) \ .
\end{equation}
This action can be mapped to the GNY action \eqref{eq: GNY action} with $\psi\to\sqrt{2}\psi$ and by identifying $y=g_1$.

Analogously, the critical $N=2$ NJLY model is expected to correspond to the Wess-Zumino model, which includes a complex scalar and a $4$-component Majorana fermion~\cite{Fei:2016sgs}. When continued to $d=3$, it becomes a $\mathcal{N}=2$ SUSY theory of a single chiral superfield in the superspace $(x^{\mu}, \theta^{\alpha}, \bar\theta^{\alpha})$ 
\begin{equation}
    \Phi(x,\theta,\bar{\theta})=\phi+\sqrt{2}\theta\psi+\theta^2 F-i\theta\gamma^{\mu}\bar\theta\partial_{\mu}\phi+\frac{i}{\sqrt{2}}\theta^2\partial_{\mu}\psi\gamma^{\mu}\bar\theta+\frac{1}{4}\theta^2\bar\theta^2\Box\phi \ ,
\end{equation}
where $\phi$ is a complex scalar and $\psi$ a 2-component Dirac fermion. Defining the covariant derivatives $D_{\alpha}=\displaystyle{\frac{\partial}{\partial\theta^{\alpha}}-i(\gamma^{\mu}\bar{\theta})_{\alpha}\partial_{\mu}}$ and $\bar{D}_{\alpha}=\displaystyle{-\frac{\partial}{\partial\bar\theta^{\alpha}}+i(\theta\gamma^{\mu})_{\alpha}\partial_{\mu}}$, we have $\bar D_{\alpha}\Phi=D_{\alpha}\bar\Phi=0$. The supersymmetric action is then
\begin{equation}
    \mathcal{S}=\int d^3x d^2\theta d^2\bar\theta \bar\Phi\Phi+\int d^3x d^2\theta W(\Phi) +\int d^3x d^2\bar\theta \bar W(\bar\Phi) \ , \qquad W(\Phi)=\frac{g_2}{3!}\Phi^3 \,.
\end{equation}
Integrating over the Grassmann coordinates $\theta$ and $\bar\theta$ and subsequently integrating $F$ out one gets
\begin{equation}
    \mathcal{L}=\partial_{\mu}\bar\phi\partial^{\mu}\phi+i\bar\psi\slashed{\partial}\psi-\frac{g_2^2}{4}(\bar\phi\phi)^2-\frac{g_2}{2}(\phi\psi\psi+h.c.) \ .
\end{equation}
This Lagrangian can be identified with the NJLY Lagrangian in eq. \eqref{eq: NJLY action} with $\phi\to\phi/\sqrt{2}$ and by identifying $y=g_2/\sqrt{2}$.

In the GNY and NJLY models, the emergent supersymmetry can be seen by continuing the fermion flavor number $N \to 1$ and $N \to 2$, respectively, and noting that the fixed-point couplings are related as $y^2 = 2 \lambda$. Moreover, from equations \eqref{eq: fermion zero mode} and \eqref{NJLYANGLE} we observe that, to the computed order, the $\ell=0$ fermionic stability angles satisfy
\begin{align}\label{eq: fermion zero mode susy}
    \nu_{F,0} = \frac{\mathcal{T}}{2} \ .
\end{align}
Therefore, exciting a fermion $\ell=0$ mode simply adds $1/2$ to the scaling dimension. For example
\begin{align}
    \delta_{F,0} : \quad \Delta_\sigma \to \Delta_\psi = \Delta_\sigma + \frac{1}{2} \ ,
\end{align}
which is the expected supersymmetric relation for the lowest superconformal multiplet. As we shall see, eq.~\eqref{eq: fermion zero mode susy} is exact.

\subsection{Associated Lamé fluctuations}
In the SUSY limit, eq.~\eqref{eq: spinor flucutation} simplifies to
\begin{align}
    \label{eq: susy spinor flucutation}
    F^{(\pm)}(\ell) &=  -\partial_u^2 - 4 m \ \cn^2(u|m) \pm i 2 \sqrt{m} \ \sn(u|m) \dn(u|m) - \Lambda_{F}(\ell) \ , \\
    \Lambda_{F}(\ell) &= (1-2m) \left(1+\frac{1+2\ell}{2 \mu}\right)^2 \ .
\end{align}
Employing the basic identities reviewed in App.~\ref{sec: app_elliptics}, we can rewrite the above as
\begin{align}\label{eq: our DTV operator}
	F^{(\pm)} = - \partial_{\tilde{u}}^2 + 6 \tilde{m} \sn^2(\tilde{u}| \tilde{m}) + 2 \tilde{m} \cd^2(\tilde{u}|\tilde{m}) - \tilde{\Lambda}_F(\ell) \ ,
\end{align}
where
\begin{align}
	\tilde{\Lambda}_F(\ell) &=  4 \tilde{m} + 2(2-\tilde{m}) \left(1+\frac{1+2\ell}{2 \mu}\right)^2 \ , \\
    \label{eq: complex modulus}
	\tilde{m} &= \frac{4 i\sqrt{\frac{m}{1-m}}}{\left(1+i \sqrt{\frac{m}{1-m}}\right)^2}  \ , \\
	\tilde{u}	&= \frac{\sqrt{1-m}+i\sqrt{m}}{2}  \left(u \pm  \mathbb{K}(m)\right)  \ .
\end{align}
The operator eq.~\eqref{eq: our DTV operator} is of the form of an associated Lamé operator\footnote{In the rest of the section, we generically denote the elliptic parameter as $m$. In the case of the fermionic fluctuations, this corresponds to $\tilde{m}$ in eq. \eqref{eq: complex modulus}.}
\begin{align} \label{eq: associated Lame}
    \text{L} = -\partial_{u}^2 + n_0(n_0+1) m \sn^2(u|m) + n_1(n_1+1) m \cd^2(u|m) - \Lambda \ .
\end{align}
with $n_0=2$ and $n_1=1$. Using similar manipulations, the standard Lamé operators with $n_1=0$ can be mapped into associated Lamé operators with $n_0 = n_1$~\cite{Ganguly:2006mx}.  For generic integer values of $n_0$ and $n_1$, this is a finite gap operator whose spectrum consists of a finite number of bound bands followed by a continuum band.
Equation \eqref{eq: associated Lame} is of the Hill type. 

In our case, with $(n_0,n_1) = (2,1)$ there are three\footnote{While a naive analysis might suggest the presence of three gaps, one has degenerate edges and is therefore closed \cite{Khare:2001dc}.} allowed bands $(\Lambda_0, \Lambda_1) , \ (\Lambda_2, \Lambda_3), \ (\Lambda_4, \infty)$. The band edges read \cite{Khare:2001dc}
\begin{align}
    \Lambda_0 &= 4m \ , \\
    \Lambda_1 &= 5+ m - 2 \sqrt{4-3m} \ , \\
    \Lambda_2 &= 5 + 2m - 2 \sqrt{4-5m+m^2} \ , \\
    \Lambda_3 &= 5+2m+2 \sqrt{4-5m+m^2} \ , \\
    \Lambda_4 &= 5+m+2 \sqrt{4-3m} \ .
\end{align}
Both the fermionic and bosonic fluctuation operators therefore have two finite gaps in the supersymmetric limit. In the fermionic case, however, the elliptic parameter $\tilde{m}$ is complex. The expressions above are therefore understood by analytic continuation and do not directly determine the stability properties of the fermionic fluctuations.

We proceed to compute the exact stability angles for the more general Darboux-Treibich-Verdier (DTV) operator, a finite-gap Schrödinger operator with an elliptic potential dependent on four integers $(n_0,n_1,n_2,n_3)$. 

\subsubsection{Stability angles of DTV operators}

The DTV operator is 
\begin{align}\label{eq: general DTV operator}
    \text{D} = - \partial_u^2 + n_0(n_0+1) \ m \ \sn^2 + n_1(n_1+1) \ m \ \frac{\cn^2}{\dn^2} + n_2(n_2+1) \frac{\dn^2}{\cn^2} + n_3(n_3+1) \frac{1}{\sn^2} - \Lambda \ .
\end{align}
To determine its stability angles, we will use the general solution first published in~\cite{deSparre:1883} $(p. \ 289-321)$ and further elucidated in~\cite{Matveev:2006dtv}. 
The two linearly independent solutions to
\begin{align}
    \text{D}\xi_\pm = 0 \ ,
\end{align}
are
\begin{align}\label{eq: DTV solution}
    \xi_\pm &= A_\pm \frac{H(u \pm a_1) \ldots H(u \pm a_\beta)}{H^{n_3}(u) H^{n_2}_1(u) \Theta^{n_1}_1(u) \Theta^{n_0}(u)} e^{\mp \frac{C}{2} u } \ , \\
    \beta &= n_0 + n_1 + n_2 + n_3 \ ,
\end{align}
where $H,H_1,\Theta,\Theta_1$ are Jacobi theta functions (App.~\ref{sec: app_elliptics}). The coefficients $A_\pm$ are normalization constants and
\begin{align}
    \frac{C}{2} =  \sum_{i=1}^\beta \frac{\Theta'(a_i)}{\Theta(a_i)} \ .
\end{align}
Using

\begin{align}
    Z(z) = \frac{\Theta'(z)}{\Theta(z)} \ ,
\end{align}
and that the function multiplying the exponential in eq.~\eqref{eq: DTV solution} is $\tau(m)\equiv \tau(m;n_0,n_1,n_2,n_3)$-periodic, with $\tau(m)=2^{(n_2+n_3)\text{mod}2}2\mathbb{K}(m)$ as can be seen from eq. \eqref{eq:Theta_transf}, we obtain an analytical expression for the stability angles
\begin{align}\label{eq: DTV stability angle}
    \nu = \pm i \tau(m) \sum_{i=1}^\beta Z(a_i|m) \ .
\end{align}
The parameters $a_i$ can be obtained in the following way. Define the polynomial 
\begin{align}\label{eq: polynomial equation}
    f(r) = \sum_{i=0}^\beta \alpha_{i-1} r^{\beta-i} \ , \quad \alpha_{-1} = 1 \ .
\end{align}
Then 
\begin{align}
    f(\sn^2(a_i|m)) = 0 \ .
\end{align}
The coefficients $\alpha_i$ can be determined recursively. The algorithm used depends on the relation between $\beta$ and $n_0$. In our case, where $\beta \leq 2n_0$ they are determined as follows. The leading coefficient reads
\begin{align}
    \alpha_0 = \frac{\Lambda - (n_0+n_1)^2-m(n_0+n_2)^2}{(2n_0-1)m} \ ,
\end{align}
while the remaining coefficients are determined recursively from
\begin{align}
    \label{eq: DTV cond1}
    &m \alpha_i
\bigg[
\beta(\beta - 2i - 2)
- (i+1)(2n_0 - 2i - 1)
\bigg] \nonumber \\
&+ \alpha_{i-1}
\bigg[
-\beta(\beta - 2i)(1+m)
+ 2i\bigg(n_0+n_1-i+m(n_0+n_2-i)\bigg)
+ (2n_0-1)m\alpha_0
\bigg] \nonumber \\
&+ \alpha_{i-2}
(\beta-i+1)(2\beta-2i-2n_3+1)
- \mu_{i-1}
+ (1+m)\mu_i
- m\mu_{i+1} = 0  \ , 
\end{align}
\begin{align}
\label{eq: DTV cond2}
&2(n_1+n_2)\mu_{i-1}
- 2(n_1+n_2m)\mu_i
- m(i+1)(2n_0-2i-1)(2n_0-i)\alpha_i \nonumber \\
&+ 2\bigg[
(2n_0-1)(n_0-i)m\alpha_0
+ \beta(\beta-2i)(n_1+n_2m) \nonumber \\
&+ i(2n_0-i)\bigg((n_0-i)(1+m)+n_1+n_2m\bigg)
\bigg]\alpha_{i-1} \nonumber \\
&- (\beta-2n_0+i)(\beta-i+1)(2\beta-2n_3-2i+1)\alpha_{i-2} = 0 \ ,
\end{align}
setting $\alpha_{-3} = \alpha_{-2} = 0$ and $\mu_i=0$ for $i < 0$. {Given $\mu_0, \ldots, \mu_s$ and $\alpha_0, \ldots, \alpha_s$, we determine $\mu_{s+1}$ from eq.~\eqref{eq: DTV cond1} with $i=s$, and then $\alpha_{s+1}$ from eq.~\eqref{eq: DTV cond2} with $i=s+1$. We then repeat for $s = 0, \ldots, \beta-2$ until $\alpha_{\beta-1}$ is known.}

\noindent
Having obtained the roots
\begin{align}
    r_i = \sn^2(a_i|m) \ , \quad i = 1, \ldots \beta \ ,
\end{align}
we can solve for the parameters $a_i$
\begin{align}
    a_i = \pm \sn^{-1}(\sqrt{r_i}|m) \ .
\end{align}
The sign of $a_1$ is arbitrary and the remaining ones are chosen so that
\begin{align}
    \frac{f'(\sn^2(a_1))}{\sn^{2n_3-1}(a_1) \cn^{2n_2-1}(a_1) \dn^{2n_1-1}(a_1)} &= \ldots =  \frac{f'(\sn^2(a_i))}{\sn^{2n_3-1}(a_i) \cn^{2n_2-1}(a_i) \dn^{2n_1-1}(a_i)} \nonumber\\
    &=  \frac{f'(\sn^2(a_\beta))}{\sn^{2n_3-1}(a_\beta) \cn^{2n_2-1}(a_\beta) \dn^{2n_1-1}(a_\beta)} \ .
\end{align}
We can formally write the stability angles as
\begin{align}
    \nu = \pm i \tau(m) \sum_{i=1}^\beta Z(\pm\sn^{-1}(\sqrt{r_i}|m)|m) \ .
\end{align}
Finally we replace the nested elliptic functions via the identity~\cite{Beccaria:2010zn}
\begin{align}
    Z\left(\sn^{-1}(\sqrt{\frac{a^2}{m}}|m)|m\right) = i \sqrt{1-a^2} \sqrt{1-\frac{m}{a^2}} \left(1-\frac{\Pi(a^2|m)}{\mathbb{K}(m)}\right) \ . 
\end{align}
We proceed to use this construction and compute the semiclassical NLO contribution exactly for both the GNY and NJLY models.

\subsubsection{Exact NLO contribution}
Comparing eq.~\eqref{eq: our DTV operator} and eq.~\eqref{eq: general DTV operator}, we observe that in the SUSY limit we have $(n_0,n_1,n_2,n_3) = (2,1,0,0)$ and therefore $\beta = 3$ and $\tau(\tilde{m}) = 2 \mathbb{K}(\tilde{m})$. Carrying out this procedure we find the non-zero coefficients
\begin{align}
    \alpha_0 &= \frac{\tilde{\Lambda}_F-9-4\tilde{m}}{3 \tilde{m}} \ , \quad &\mu_0 &= 9 \ , \\
    \alpha_1 &= \frac{3+\alpha_0(2+\tilde{m}+\alpha_0)}{\tilde{m}} \ , \quad &\mu_1 &= 3 \alpha_0 \ , \\
    \alpha_2 &=- \frac{12+\alpha_0(8+(4+3\alpha_0)\tilde{m})}{3 \tilde{m}^2} \ , \quad &\mu_2 &= - \frac{(2 \alpha_0+3)(\tilde{m} \alpha_0+3)}{\tilde{m}} \ .
\end{align}
The exact expressions for the roots of eq.~\eqref{eq: polynomial equation} are lengthy and we omit them here. We find
\begin{align}
    \frac{1}{2} \sum_{\ell=1}^{\infty}g_F(\ell) \tilde{\nu}_F(\ell) = \tilde{\Omega}_F + \sum_{\ell=1}^{\infty} \tilde{\rho}_{F,\ell}
\end{align}
with
\begin{align}
    \tilde{\Omega}_F &= \pi -\frac{5 (2-\tilde{m}) \mathbb{K}(\tilde{m})+14 \mathbb{E}(\tilde{m})}{4 \sqrt{2} \sqrt{2-\tilde{m}}} \ , \\
    \tilde{\rho}_{F,\ell} &= \frac{1}{2}g_F(\ell) \tilde{\nu}_{F,\ell} +\frac{1}{2} \pi  (\ell+1) (\ell+2) \nonumber\\
    &\ \ - \frac{1}{12 \sqrt{2} \ell (2-\tilde{m})^{3/2}} \bigg[4 \left(6 \ell^2+9 \ell+2\right) (2-\tilde{m}) \mathbb{E}(\tilde{m})  \nonumber\\
    &\ \ + \left(\left(12 \ell^4+54 \ell^3+66 \ell^2+18 \ell\right) (2-\tilde{m})^2+\tilde{m} (16-3 \tilde{m})-16\right)\mathbb{K}(\tilde{m})\bigg] \ .
\end{align}
The pole term is
\begin{align}
    F_{\text{pole}}= -2 \times \frac{(4-\tilde{m}) (4-3 \tilde{m}) \mathbb{K}(\tilde{m})-8 (2-\tilde{m}) \mathbb{E}(\tilde{m})}{12 \sqrt{4-2 \tilde{m}} (2-\tilde{m}) \epsilon} \ ,
\end{align}
where the factor of $2$ in front stems from having the two $(\pm)$ fermionic determinants.  Substituting $\tilde{m}=4 i \sqrt{m/(1-m)} (1+i\sqrt{m/(1-m)})^{-2}$ (eq.~\eqref{eq: complex modulus}) and using the identities
\begin{align}
    \mathbb{K}(\tilde{m})&=\left(\sqrt{1-m}+i \sqrt{m}\right) \mathbb{K}(m) \ , \\
    \mathbb{E}(\tilde{m}) &= \left(\sqrt{1-m}-i \sqrt{m}\right) (2 \mathbb{E}(m)-\mathbb{K}(m)) \ ,
\end{align}
we obtain
\begin{align}
    F_{\text{pole}}=-2 \times \frac{(1-m) (2-3 m) \mathbb{K}(m)-(2-4 m) \mathbb{E}(m)}{3 (1-2 m)^{3/2} \epsilon} \ .
\end{align}
Combining with equations \eqref{eq: action pole} and \eqref{eq: boson pole}, one can verify that the pole correctly cancels
\begin{align}
    S_{\text{pole}}-B_{\text{pole}}+F_{\text{pole}}=0 \ .
\end{align}
An analogous cancellation occurs in the NJLY model. Finally, we have
\begin{align}
    \nu_{F,0} = \frac{\mathcal{T}}{2} \,,
\end{align}
which provides nontrivial evidence that the GNY and NJLY models flow to a superconformal theory in the IR. Let us consider some explicit examples in the GNY model. Analogous expressions appear in the NJLY model. In CFTs with supersymmetry, the conformal multiplet structure is further organized into supermultiplets where scalars and spinors in the same supermultiplet are related by transformations generated by supercharges $Q$ with dimensions $1/2$. Exciting a fermionic $\ell=0$ mode in the SUSY limit amounts to acting with a supercharge. Let us consider explicit examples in the perturbative limit. In the ancillary file we provide eight-loop order results for the complete spectrum, while here we restrict ourselves to two-loop order as in Sec.~\ref{sec:explicit_results}

\begin{itemize} 
    \item Ground state operator: 
    \begin{align}
        \Delta_{\sigma^n} &= n+ \frac{3}{14} (n-3) \eps n - \frac{1}{196} (17 n-52) (\eps n)^2 + \ldots  \ . 
    \end{align}
    \item $q_{F,\ell} = k \delta_{\ell,0}$ with $k = 1,2$: 
    \begin{align}
        k = 1 : \quad \Delta_{Q \sigma^n} &= \Delta_{\sigma^n} + \frac{1}{2} \ , \\
        k = 2 : \quad \Delta_{Q^2 \sigma^n} &= \Delta_{\sigma^n} + 1 \ ,
    \end{align}
    where we denote the composite operators according to the three-dimensional supersymmetric notation. For $n=1$, these operators correspond to $\psi$ and $F$, where the conformal descendant $\partial \sigma^n$ arising at $k=2$ from multiplying two spinor representations is removed by contracting the indices of the supercharges. The lowest dimensional multiplet is the Wess-Zumino multiplet which on shell consists of $\{\sigma, \psi, \sigma^2\}$ with dimensions related by 
\begin{align}
    \Delta_{\sigma^2} = \Delta_{\sigma} + 1 = \Delta_{\psi}+\frac{1}{2} \ ,
\end{align}
This gives the hyperscaling relations
\begin{align}
    \eta_\sigma = \eta_\psi \ , \quad \frac{1}{\nu} = \frac{d-\eta_{\sigma}}{2} \ .
\end{align}
Here $\eta_\sigma$ and $\eta_\psi$ are  critical exponents defined by $\Delta_\sigma=(d-2+\eta_\sigma)/2$ and $\Delta_\psi=(d-1+\eta_\psi)/2$.
The correlation-length exponent $\nu$ is related to the scaling dimension of the $\sigma^2$ operator by $\nu^{-1}=d-\Delta_{\sigma^2}$.

From the EOM of $\psi$ we find that $\Delta_{\sigma \psi} = \Delta_{\psi} + 1$. Combining with the SUSY relations we obtain the supersymmetric relation $\Delta_{\sigma \psi} = \Delta_{\sigma^2} + \frac{1}{2}$. This, together with $\Delta_{\psi} = \Delta_{\sigma} + \frac{1}{2}$ are explicit algebraic examples of the assertion that exciting a single $\ell=0$ fermionic mode has the effect $\delta_{F,0}: \ \sigma^{n} \to \sigma^{n-1} \psi$.

    \item  $q_{F,\ell} = \delta_{\ell,s\ge1}$: 
    \begin{align}
        \Delta_{\sigma^{n-1}\partial^s \psi } &= n+s+\frac{1}{2}+\frac{1}{14}  \left(3 n+\frac{4}{s+1}+\frac{4}{s+2}-15\right) \eps n  \nonumber\\
        &\ \ + \frac{1}{196 (1 + s)^3 (2 + s)^3} \bigg[440 - 17 n (1 + s)^3 (2 + s)^3 \nonumber\\
        &\ \ + 
 s (2480 + s (5178 + s (5409 + s (3039 + s (879 + 103 s)))))\bigg](\eps n)^2 + \ldots \ .
    \end{align}
\end{itemize}
Having presented the exact solution and perturbative results in the limit of emergent supersymmetry, in the next section we discuss the large $\eps n$ regime.

\subsection{Large $\eps n$ regime}

The discussion of the large \(\epsilon n\) regime parallels that of the \(O(N)\) model presented in \cite{Antipin:2025rsr}. At leading semiclassical order, the scaling dimensions behave as \(\Delta_{n,q_\ell}\sim\epsilon^{1/3}n^{4/3}\). At next-to-leading order, however, some bosonic stability angles become complex for sufficiently large \(\epsilon n\). Their imaginary parts are related to the Lyapunov exponents of the linearized fluctuation equations and signal instabilities of the classical saddle.

In the GNY model, each \(\ell\geq2\) has an instability interval in which \(\nu_{B,\ell}\) becomes complex because the spectral parameter \(\Lambda_B(\ell)\) in eq.~\eqref{Lambada} lies in the upper finite gap of the two-gap Lamé operator \cite{Antipin:2025rsr}. Both endpoints of these intervals increase with \(\ell\). For moderate \(\epsilon n\), the intervals are disjoint and separated by stable regions that cannot be reached from the free theory along the real coupling axis without crossing an unstable region. At sufficiently large \(\epsilon n\), successive intervals overlap, so that at least one stability angle is always complex. For \(N=1\), the first instability occurs in the \(\ell=2\) mode for \(4.4\lesssim\epsilon n\lesssim5.1\).

The longitudinal modes \(\nu_{\parallel,\ell}\) of the NJLY model exhibit analogous behaviour, the only difference being the fixed point value of the quartic coupling. In particular, for $N=2$, $\nu_{\parallel,\ell=2}$ is complex for $3.8 \lesssim  \eps n \lesssim 4.4$. On the other hand, the transverse modes behave differently: for each \(\ell\geq1\), there is a threshold above which \(\nu_{\perp,\ell}\) remains complex because the spectral parameter lies in the gap region of the one-gap Lamé operator \cite{Antipin:2025rsr}.

The fate of the semiclassical expansion in the large $\eps n$ regime remains an interesting open question, which we leave for future work.

\section{Conclusions}
 
We have determined the spectrum of scaling dimensions of Chiral Ising and XY  CFTs using a semiclassical methodology. For each theory, the path integral is expanded around a classical time-periodic homogeneous configuration and quantized according to the Bohr-Sommerfeld condition. In the double scaling limit of small coupling and large operator dimension, we achieve a systematic and controllable semiclassical expansion. Adding scalar and fermionic quantum fluctuations on top of the classical background, we have been able to disentangle the Lorentz structure of the composite operators.  We have compared our results to existing literature and provided new ones for a wider class of composite operators.

In the limit of emergent supersymmetry, the fermionic fluctuations also take the form of a finite-gap associated Lamé operator, an example of the general Darboux-Treibich-Verdier operator. This has allowed us to employ a general procedure for computing the associated stability angles. The latter has been presented in a way to make it easily applicable to future semiclassical analyses across several realms of physics. Finally, our results provide further evidence for emergent supersymmetry by independently determining scaling relations to infinite order in perturbation theory in the semiclassical limit.

\section*{Acknowledgments}
{The work of J.H., G.M. and F.S. is partially supported by the Carlsberg Foundation, semper ardens grant CF22-0922. The work of J.B. is supported by the Swiss National Science Foundation under grant number 200021\_219267.}

\newpage
\appendix

\section{Elliptic functions}\label{sec: app_elliptics}
In this section we summarize a few useful elliptic function and integral properties that were used in carrying out our computations. All the integrals and functions depend on a real number $k\in(0,1)$ called the modulus and its complementary $k'=\sqrt{1-k^2}$. In the main text we used $m$ related as $m=k^2$. Firstly, the \emph{incomplete elliptic integrals of the first, second, and third kind} are defined as
\begin{align}
    F(\varphi|k^2) &\equiv  \int_0^{\sin \varphi} \frac{d t}{\sqrt{(1-k^2t^2)(1-t^2)}},\\
    E(\varphi|k^2) &\equiv \int_0^{\sin \varphi} dt ~ \sqrt{\frac{1-k^2t^2}{1-t^2}},\\
    \Pi(n|\varphi,k^2) &\equiv \int_0^{\sin \varphi} \frac{dt}{(1-nt^2) \sqrt{(1-t^2)(1-k^2t^2)}} ,
\end{align}
with the corresponding \textit{complete} elliptic integrals found by setting $\varphi=\pi/2$: 
\be
\mathbb{K}(k^2) \equiv F(\pi/2|k^2)\,, \quad \ \ELE(k^2) \equiv E(\pi/2|k^2)\,, \quad \ \Pi(n|k^2) \equiv\Pi(n|\pi/2,k^2)
\ee
Useful identities are
\begin{align}
    \mathbb{K}(\tilde{k}^2)&=\left(k'+i k\right) \mathbb{K}(k^2) \ , \\
    \mathbb{E}(\tilde{k}^2) &= \left(k'-i k\right) (2 \mathbb{E}(k^2)-\mathbb{K}(k^2)) \ ,
\end{align}
where
\begin{align}
    \tilde{k}^2 = \frac{4i k k'}{(k'+ik)^2} \ \Rightarrow \ \tilde{k}'^2 = \left(\frac{k'-ik}{k'+ik}\right)^2 \ .
\end{align}
Alternatively
\begin{align}\label{eq: KEEK identity}
	\mathbb{K}(k^2) &= \frac{1+\tilde{k}'}{2} \mathbb{K}(\tilde{k}^2)\ , \quad 
	\mathbb{E}(k^2) = \frac{1}{1+\tilde{k}'}\left(\mathbb{E}(\tilde{k}^2) + \tilde{k}' \mathbb{K}(\tilde{k}^2) \right) \ ,
\end{align}
where
\begin{align}
	\tilde{k}^2 = \frac{4k}{(1+k)^2} \ \Rightarrow \ \tilde{k}'^2 = \left(\frac{1-k}{1+k}\right)^2 \ .
\end{align}

Secondly, we introduce the \emph{Jacobi elliptic functions} $\sn,\cn,\dn$  defined by
\be
\cn(u\,|\,k^2)=\cos \varphi,~~~~~~~\sn(u\,|\,k^2)=\sin \varphi,~~~~~~\dn(u\,|\,k^2)=(1-k^2\,\sin^2 \varphi)^{1/2},
\ee 
where $\varphi$ is the \emph{Jacobi amplitude} defined as the inverse of the incomplete elliptic integral of the first kind, i.e., as the solution of $F(\varphi|k^2) = u$
\be
\varphi={\rm am} (u\,|\,k^2),~~~~{\rm with}~~~~u=\int_0^\varphi d\theta\  (1-k^2 \,\sin^2\theta)^{-1/2}.
\ee
The elliptic functions are doubly periodic and their period lies in the complex plane of $u$. Each has a real period and a complex one. We report here the real ones, which are $2\KK$ for $\dn$ and $4\KK$ for $\sn$ and $\cn$. It is worth mentioning the quarter-period argument shifts
\begin{align}
	\sn(u \pm \mathbb{K}|k^2) &= \pm \cd(u|k^2) \ , \\
	\cn(u \pm \mathbb{K}|k^2) &= \mp k' \sd(u|k^2) \ , \\
	\dn(u \pm \mathbb{K}|k^2) &= k' \nd(u|k^2) \ .
\end{align}
The two transformations used for the modulus parameter are the imaginary modulus transformation
\begin{align}
    \sn\Bigg(\frac{1}{k'}u|k^2\Bigg) &= \frac{1}{k'} \sd\Bigg(u|\frac{-k^2}{k'^2}\Bigg) \ , \\
    \cn\Bigg(\frac{1}{k'}u|k^2\Bigg) &= \cd\Bigg(u|\frac{-k^2}{k'^2}\Bigg) \ , \\
    \dn\Bigg(\frac{1}{k'}u|k^2\Bigg) &= \nd\Bigg(u|\frac{-k^2}{k'^2}\Bigg) \ .
\end{align}
and the Gauss transformation
\begin{align}
\sn((1+\tilde{k})u | k^2) 
&= \frac{(1+\tilde{k})\,\sn(u | \tilde{k}^2)}
{1+\tilde{k}\,\sn^2(u | \tilde{k}^2)} \ , \\
\cn((1+\tilde{k})u | k^2) 
&= \frac{\cn(u | \tilde{k}^2)\dn(u | \tilde{k}^2)}
{1+\tilde{k}\,\sn^2(u | \tilde{k}^2)} \ , \\
\dn((1+\tilde{k})u | k^2) 
&= \frac{1-\tilde{k}\,\sn^2(u | \tilde{k}^2)}
{1+\tilde{k}\,\sn^2(u | \tilde{k}^2)} \ ,
\end{align}
with
\begin{align}
k^2 = \frac{4\tilde{k}}{(1+\tilde{k})^2}, 
\qquad
k'^2 = \left(\frac{1-\tilde{k}}{1+\tilde{k}} \right)^2.
\end{align}
The relevant Jacobi theta functions are
\begin{align}
    \Theta(u) &= \vartheta_4\Bigg(\frac{u}{\vartheta_3(0,q)^2},q\Bigg) \ , \\
    \Theta_1(u) &= \vartheta_3\Bigg(\frac{u}{\vartheta_3(0,q)^2},q\Bigg) \ , \\
    H(u) &= \vartheta_1\Bigg(\frac{u}{\vartheta_3(0,q)^2},q\Bigg) \ , \\
    H_1(u) &= \vartheta_2\Bigg(\frac{u}{\vartheta_3(0,q)^2},q\Bigg) \ ,
\end{align}
where
\begin{align}
    q = e^{-\frac{\pi  \mathbb{K}\left(k'^2\right)}{\mathbb{K}\left(k^2\right)}} \ .
\end{align}
Finally, some transformation properties of Jacobi theta functions are
\begin{equation}
\label{eq:Theta_transf}
    \begin{split}
        H(x+K)&=H_1(x) \ ,\\
        H_1(x+K)&=-H(x) \ ,\\
        \Theta(x+K)&=\Theta_1(x) \ ,\\
          \Theta_1(x+K)&=\Theta(x) \ .
    \end{split}
\end{equation}

\section{Renormalization} \label{sec: renormalize}
In this section, we illustrate the renormalization procedure in the GNY case. In order to renormalize the classical action, we evaluate it in terms of bare quantities
\begin{align}
    S_\text{cl} &= s(m) \frac{(d-2)^2 \pi ^{d/2}}{\lambda_0 \Gamma \left(\frac{d}{2}-1\right)} \ , \\
    s(m) &= \frac{2 ((m-1) (3 m-2) \mathbb{K}(m)+(4 m-2) \mathbb{E}(m))}{3 (1-2 m)^{3/2}} \ .
\end{align}
The bare coupling in terms of the renormalized ones is
\begin{align}
    \lambda_0 = M^{\epsilon } \left(\lambda +\frac{9 \lambda ^2-N y^4+\lambda  N y^2 }{8 \pi ^2 \epsilon }\right) \ .
\end{align}
Here the coefficient of the $\epsilon$-pole is the one-loop term in the beta function of $\lambda$. Then
\begin{align}
    S_\text{cl} = s(m) \frac{\pi ^{2-\frac{\epsilon }{2}} (\epsilon -2)^2 M^{-\epsilon }}{\Gamma \left(1-\frac{\epsilon }{2}\right)} \left(\frac{1}{\lambda }+\frac{-9 \lambda ^2+N y^4-\lambda  N y^2}{8 \pi ^2 \lambda ^2 \epsilon }\right) \ .
\end{align}
The first term is the classical action expressed in terms of the renormalized coupling. The second term is a quantum correction. In the limit $\epsilon \to 0$ it contributes a divergent term
\begin{align}\label{eq: action pole}
    S_{\text{pole}} = \frac{((m-1) (3 m-2) \mathbb{K}(m)+(4 m-2) \mathbb{E}(m)) \left(N \left(y^4-\lambda  y^2\right)-9 \lambda ^2\right)}{3 \lambda ^2 (1-2 m)^{3/2} \epsilon } \ ,
\end{align}
which, as we shall see below, will cancel the divergence stemming from the sum over stability angles. The finite regular part is
\begin{align}
    S_{reg} = s(m) \bigg(\frac{\pi ^{2-\frac{\epsilon }{2}} (\epsilon -2)^2 M^{-\epsilon }}{\Gamma \left(1-\frac{\epsilon }{2}\right)} \frac{1}{\lambda} + \frac{(2 \log (M)+\gamma +2+\log (\pi )) \left(9 \lambda ^2-N y^4+\lambda  N y^2\right)}{4 \lambda ^2}\bigg) \ .
\end{align}
By computing the contribution to the energy according to~\cite{Antipin:2025rsr}
\begin{align}\label{eq: timederivofaction}
    -\frac{\partial S_{reg}}{\partial \T}\bigg|_{cl} = E_{cl} + \delta E_1 + \ldots \ ,
\end{align}
and finally taking the $\epsilon \to 0$ limit, we obtain 
\begin{align}
    \delta E_1 = - nC_0(\lambda n) \frac{9 \lambda  (N+6)}{8 \pi ^2 \left(-N+\sqrt{N (N+132)+36}+6\right)} \ .
\end{align}
where the dots refer to higher-order semiclassical corrections.

\section{Perturbative computation of stability angles}\label{sec: perturbative}
In this section we describe the perturbative computation of stability angles used to calculate the functional determinant of the fermionic operators perturbatively. To be explicit let us consider the $(-)$ case in eq.~\eqref{eq: spinor flucutation}
\begin{align}
    \label{eq: spinor flucutation repeat}
    F_j^{-}(\ell) &=  -\partial_u^2 - 2 m \frac{y^2} {\lambda} \ \cn^2(u|m) - i \sqrt{2m} \frac{y}{\sqrt{\lambda}} \ \sn(u|m) \dn(u|m) - \Lambda_{F}(\ell)  \ , \\
    \Lambda_{F}(\ell) &= (1-2m)\left(1+\frac{1+2\ell}{2 \mu}\right)^2  \ .
\end{align}
In practice we will always change coordinates
\begin{equation}
    z=\frac{2\pi}{4\mathbb{K}(m)}u
\end{equation}
so that the potential is $ 2\pi$-periodic in $z$. We solve
\begin{align}
    F_j^{(-)}(\ell) \xi_\ell(z) = 0 \ ,
\end{align}
with an ansatz
\begin{align}
    \xi_\ell(z) = e^{i z (k_0 + \sqrt{m} k_{1} + m k_2 + \ldots)} \left(\chi_0(z) +\sqrt{m}\chi_{1}(z) + m \chi_2(z) + \ldots \right) \ ,
\end{align}
which can be solved for the $\chi$'s as an expansion in $\sqrt{m}$. At each order $i$ we impose that $\chi_i$ is periodic, which in turn determines $k_i$. The integration constants are always chosen so that a Bloch solution is found, i.e., $k_i$ is independent of $z$. The $\ell=0$ mode must be computed separately by substituting directly $\ell=0$ into the operator eq.~\eqref{eq: spinor flucutation repeat}. The stability angles are then given by
\begin{align}
    \nu_{F,\ell} = 2 \pi (k_0 +\sqrt{m} k_1 + m k_2 + \ldots 
    ) \ .
\end{align}
Finally, we choose the sign and shift the stability angles by multiples of $2 \pi$ so that $\nu_{F,\ell}\geq 0$.


\newpage

\begin{thebibliography}{1}
\bibitem{Fei:2016sgs}
L.~Fei, S.~Giombi, I.~R.~Klebanov and G.~Tarnopolsky,
``Yukawa CFTs and emergent supersymmetry,''
\textit{PTEP} \textbf{2016} (2016) no.12, 12C105
doi:10.1093/ptep/ptw120
[arXiv:1607.05316 [hep-th]].

\bibitem{Gross:1974jv}
D.~J.~Gross and A.~Neveu,
``Dynamical symmetry breaking in asymptotically free field theories,''
\textit{Phys. Rev. D} \textbf{10} (1974), 3235--3253
doi:10.1103/PhysRevD.10.3235

\bibitem{Nambu:1961tp}
Y.~Nambu and G.~Jona-Lasinio,
``Dynamical model of elementary particles based on an analogy with superconductivity. I,''
\textit{Phys. Rev.} \textbf{122} (1961), 345--358
doi:10.1103/PhysRev.122.345



\bibitem{Herbut:2009qb}
I.~F.~Herbut, V.~Juričić and B.~Roy,
``Theory of interacting electrons on the honeycomb lattice,''
\textit{Phys. Rev. B} \textbf{79} (2009), 085116
doi:10.1103/PhysRevB.79.085116
[arXiv:0811.0610 [cond-mat.str-el]].

\bibitem{Hasenfratz:1991it}
A.~Hasenfratz, P.~Hasenfratz, K.~Jansen, J.~Kuti, and Y.~Shen,
``The Equivalence of the Top Quark Condensate and the Elementary Higgs Field,''
Nucl.\ Phys.\ B \textbf{365}, 79--97 (1991),
\href{https://doi.org/10.1016/0550-3213(91)90607-Y}{doi:10.1016/0550-3213(91)90607-Y}.

\bibitem{Zinn-Justin:1991yn}
J.~Zinn-Justin,
``Four-Fermion Interaction Near Four Dimensions,''
Nucl.\ Phys.\ B \textbf{367}, 105--122 (1991),
\href{https://doi.org/10.1016/0550-3213(91)90043-W}{doi:10.1016/0550-3213(91)90043-W}.


\bibitem{Herbut:2006cs}
I.~F.~Herbut,
``Interactions and phase transitions on graphene's honeycomb lattice,''
\textit{Phys. Rev. Lett.} \textbf{97} (2006), 146401
doi:10.1103/PhysRevLett.97.146401
[arXiv:cond-mat/0606195].

\bibitem{Herbut:2009vu}
I.~F.~Herbut, V.~Juričić and O.~Vafek,
``Relativistic Mott criticality in graphene,''
\textit{Phys. Rev. B} \textbf{80} (2009), 075432
doi:10.1103/PhysRevB.80.075432
[arXiv:0904.1019 [cond-mat.str-el]].


\bibitem{Louvet_2016}
T.~Louvet, D.~Carpentier and A.~A.~Fedorenko,
``On the disorder-driven quantum transition in three-dimensional relativistic metals,''
\textit{Phys. Rev. B} \textbf{94} (2016) no.22, 220201(R)
doi:10.1103/PhysRevB.94.220201

\bibitem{Roy_2013}
B.~Roy, V.~Juričić and I.~F.~Herbut,
``Quantum superconducting criticality in graphene and topological insulators,''
\textit{Phys. Rev. B} \textbf{87} (2013) no.4, 041401(R)
doi:10.1103/PhysRevB.87.041401


\bibitem{Zerf_2017}
N.~Zerf, L.~N.~Mihaila, P.~Marquard, I.~F.~Herbut and M.~M.~Scherer,
``Four-loop critical exponents for the Gross-Neveu-Yukawa models,''
\textit{Phys. Rev. D} \textbf{96} (2017) no.9, 096010
doi:10.1103/PhysRevD.96.096010

\bibitem{Thomas:2005}
S.~Thomas,
``Emergent Supersymmetry,''
talk at the KITP Conference \textit{Quantum Phase Transitions},
Kavli Institute for Theoretical Physics, January 21, 2005,
\url{https://online.kitp.ucsb.edu/online/qpt-c05/thomas/}.

\bibitem{Grover:2013rc}
T.~Grover, D.~N.~Sheng and A.~Vishwanath,
``Emergent space-time supersymmetry at the boundary of a topological phase,''
\textit{Science} \textbf{344} (2014) no.6181, 280--283
doi:10.1126/science.1248253
[arXiv:1301.7449 [cond-mat.str-el]].

\bibitem{Ponte:2012ru}
P.~Ponte and S.-S.~Lee,
``Emergence of supersymmetry on the surface of three dimensional topological insulators,''
\textit{New J. Phys.} \textbf{16} (2014) no.1, 013044
doi:10.1088/1367-2630/16/1/013044
[arXiv:1206.2340 [cond-mat.str-el]].


\bibitem{Gracey:2025aoj}
J.~A.~Gracey, A.~Maier, P.~Marquard and Y.~Schröder,
``Anomalous dimensions and critical exponents for the Gross-Neveu-Yukawa model at five loops,''
\textit{Phys. Rev. D} \textbf{112} (2025) no.8, 085029
doi:10.1103/lmpx-n3mj
[arXiv:2507.22594 [hep-th]].

\bibitem{Janssen:2014gea}
L.~Janssen and I.~F.~Herbut,
``Antiferromagnetic critical point on graphene's honeycomb lattice: A functional renormalization group approach,''
\textit{Phys. Rev. B} \textbf{89} (2014), 205403
doi:10.1103/PhysRevB.89.205403
[arXiv:1402.6277 [cond-mat.str-el]].
Addendum: \textit{Phys. Rev. B} \textbf{102} (2020), 199902.

\bibitem{Vacca:2015nta}
G.~P.~Vacca and L.~Zambelli,
``Multimeson Yukawa interactions at criticality,''
\textit{Phys. Rev. D} \textbf{91} (2015) no.12, 125003
doi:10.1103/PhysRevD.91.125003
[arXiv:1503.09136 [hep-th]].



\bibitem{Manashov:2017rrx}
A.~N.~Manashov and M.~Strohmaier,
``Correction exponents in the Gross--Neveu--Yukawa model at $1/N^2$,''
\textit{Eur. Phys. J. C} \textbf{78} (2018) no.6, 454
doi:10.1140/epjc/s10052-018-5902-1
[arXiv:1711.02493 [hep-th]].

\bibitem{Gracey:2017fzu}
J.~A.~Gracey,
``Critical exponent $\omega$ in the Gross-Neveu-Yukawa model at $O(1/N)$,''
\textit{Phys. Rev. D} \textbf{96} (2017) no.6, 065015
doi:10.1103/PhysRevD.96.065015
[arXiv:1707.05275 [hep-th]].

\bibitem{Gracey:1993kc}
J.~A.~Gracey,
``Computation of critical exponent $\eta$ at $O(1/N^3)$ in the four-Fermi model in arbitrary dimensions,''
\textit{Int. J. Mod. Phys. A} \textbf{9} (1994), 727--744
doi:10.1142/S0217751X94000340
[arXiv:hep-th/9306107].

\bibitem{Bashkirov:2013vya}
D.~Bashkirov,
``Bootstrapping the $\mathcal{N}=1$ SCFT in three dimensions,''
[arXiv:1310.8255 [hep-th]].

\bibitem{Iliesiu:2015qra}
L.~Iliesiu, F.~Kos, D.~Poland, S.~S.~Pufu, D.~Simmons-Duffin and R.~Yacoby,
``Bootstrapping 3D fermions,''
\textit{JHEP} \textbf{03} (2016), 120
doi:10.1007/JHEP03(2016)120
[arXiv:1508.00012 [hep-th]].

\bibitem{Iliesiu:2017nrv}
L.~Iliesiu, F.~Kos, D.~Poland, S.~S.~Pufu and D.~Simmons-Duffin,
``Bootstrapping 3D fermions with global symmetries,''
\textit{JHEP} \textbf{01} (2018), 036
doi:10.1007/JHEP01(2018)036
[arXiv:1705.03484 [hep-th]].

\bibitem{Atanasov:2022bpi}
A.~Atanasov, A.~Hillman, D.~Poland, J.~Rong and N.~Su,
``Precision bootstrap for the $\mathcal{N}=1$ super-Ising model,''
\textit{JHEP} \textbf{08} (2022), 136
doi:10.1007/JHEP08(2022)136
[arXiv:2201.02206 [hep-th]].

\bibitem{Rong:2018okz}
J.~Rong and N.~Su,
``Bootstrapping the minimal $\mathcal{N}=1$ superconformal field theory in three dimensions,''
\textit{JHEP} \textbf{06} (2021), 154
doi:10.1007/JHEP06(2021)154
[arXiv:1807.04434 [hep-th]].





\bibitem{Atanasov:2018kqw}
A.~Atanasov, A.~Hillman and D.~Poland,
``Bootstrapping the minimal 3D SCFT,''
\textit{JHEP} \textbf{11} (2018), 140
doi:10.1007/JHEP11(2018)140
[arXiv:1807.05702 [hep-th]].

\bibitem{Antipin:2022naw}
O.~Antipin, J.~Bersini and P.~Panopoulos,
``Yukawa interactions at large charge,''
\textit{JHEP} \textbf{10} (2022), 183
doi:10.1007/JHEP10(2022)183
[arXiv:2208.05839 [hep-th]].

\bibitem{Bersini:2025lxs}
J.~Bersini, S.~Hellerman, D.~Orlando and S.~Reffert,
``Resurgence analysis of the Nambu-Jona-Lasinio model at large charge,''
\textit{JHEP} \textbf{09} (2025), 159
doi:10.1007/JHEP09(2025)159
[arXiv:2505.21631 [hep-th]].


\bibitem{Chandrasekharan:2013aya}
S.~Chandrasekharan and A.~Li,
``Quantum critical behavior in three dimensional lattice Gross-Neveu models,''
\textit{Phys. Rev. D} \textbf{88} (2013), 021701
doi:10.1103/PhysRevD.88.021701
[arXiv:1304.7761 [hep-lat]].


\bibitem{Wang:2014cbw}
L.~Wang, P.~Corboz and M.~Troyer,
``Fermionic quantum critical point of spinless fermions on a honeycomb lattice,''
\textit{New J. Phys.} \textbf{16} (2014) no.10, 103008
doi:10.1088/1367-2630/16/10/103008
[arXiv:1407.0029 [cond-mat.str-el]].

\bibitem{Hands:1992be}
S.~Hands, A.~Kocic and J.~B.~Kogut,
``Four Fermi theories in fewer than four dimensions,''
\textit{Annals Phys.} \textbf{224} (1993), 29--89
doi:10.1006/aphy.1993.1039
[arXiv:hep-lat/9208022].


\bibitem{Tang:2025wtj}
Y.~Tang, C.~Voinea, L.~Hu, Z.~Papić and W.~Zhu,
``Emergence of 3D superconformal Ising criticality on the fuzzy sphere,''
\textit{Phys. Rev. Lett.} \textbf{137} (2026) no.2, 026502
doi:10.1103/zxtq-d3mk
[arXiv:2512.25054 [cond-mat.str-el]].

\bibitem{Zhou:2026ugg}
Z.~Zhou and Y.-C.~He,
``Advancing fuzzy-sphere CFTs: $\mathrm{SO}(3)$-rotation-resolving exact diagonalization, 3D Ising heavy primaries, and quasi-hole-space projection,''
[arXiv:2609.35960 [cond-mat.str-el]].

\bibitem{Antipin:2024ekk}
O.~Antipin, J.~Bersini and F.~Sannino,
``Exact results for scaling dimensions of neutral operators in scalar conformal field theories,''
\textit{Phys. Rev. D} \textbf{111} (2025) no.4, L041701
doi:10.1103/PhysRevD.111.L041701
[arXiv:2408.01414 [hep-th]].


\bibitem{Antipin:2025nlq}
O.~Antipin, J.~Bersini, J.~Hafjall, G.~Muco and F.~Sannino,
``Semiclassical spectrum of the Ising CFT,''
Phys. Rev. D \textbf{114} (2026) no.6, 6
doi:10.1103/43ck-2h6m
[arXiv:2511.08276 [hep-th]].

\bibitem{Antipin:2025rsr}
O.~Antipin, J.~Bersini, J.~Hafjall, G.~Muco and F.~Sannino,
``Semiclassical canovaccio for composite operators,''
\textit{JHEP} \textbf{05} (2026), 137
doi:10.1007/JHEP05(2026)137
[arXiv:2512.23539 [hep-th]]

\bibitem{Beccaria:2010ry}
M.~Beccaria, G.~V.~Dunne, V.~Forini, M.~Pawellek and A.~A.~Tseytlin,
``Exact computation of one-loop correction to energy of spinning folded string in $AdS_5 \times S^5$,''
\textit{J. Phys. A} \textbf{43} (2010), 165402
doi:10.1088/1751-8113/43/16/165402
[arXiv:1001.4018 [hep-th]].

\bibitem{Greene:1997fu}
P.~B.~Greene, L.~Kofman, A.~D.~Linde and A.~A.~Starobinsky,
``Structure of resonance in preheating after inflation,''
\textit{Phys. Rev. D} \textbf{56} (1997), 6175--6192
doi:10.1103/PhysRevD.56.6175
[arXiv:hep-ph/9705347].



\bibitem{Togawa:2016}
Y.~Togawa, Y.~Kousaka, K.~Inoue and J.-i.~Kishine,
``Symmetry, structure, and dynamics of monoaxial chiral magnets,''
\textit{J. Phys. Soc. Jpn.} \textbf{85} (2016) no.11, 112001

\bibitem{Erramilli:2022kgp}
R.~S.~Erramilli, L.~V.~Iliesiu, P.~Kravchuk, A.~Liu, D.~Poland and D.~Simmons-Duffin,
``The Gross-Neveu-Yukawa archipelago,''
\textit{JHEP} \textbf{02} (2023), 036
doi:10.1007/JHEP02(2023)036
[arXiv:2210.02492 [hep-th]].

\bibitem{Dondi:2022zna}
N.~Dondi, S.~Hellerman, I.~Kalogerakis, R.~Moser, D.~Orlando and S.~Reffert,
``Fermionic CFTs at large charge and large $N$,''
\textit{JHEP} \textbf{08} (2023), 180
doi:10.1007/JHEP08(2023)180
[arXiv:2211.15318 [hep-th]].

\bibitem{Osborn:1993cr}
H.~Osborn and A.~C.~Petkou,
``Implications of conformal invariance in field theories for general dimensions,''
\textit{Annals Phys.} \textbf{231} (1994), 311--362
doi:10.1006/aphy.1994.1045
[arXiv:hep-th/9307010].


\bibitem{Sannino:2026ena}
F.~Sannino,
``Lectures on Semiclassical Methods for Composite Operators,''
[arXiv:2606.11317 [hep-th]]. To appear as a Springer Lecture Note. 

\bibitem{dhn}
R.~F.~Dashen, B.~Hasslacher and A.~Neveu,
``The Particle Spectrum in Model Field Theories from Semiclassical Functional Integral Techniques,''
Phys. Rev. D \textbf{11} (1975), 3424
doi:10.1103/PhysRevD.11.3424



\bibitem{book1}
E.~T.~Whittaker, G.~N.~Watson,
``A Course of Modern Analysis,''
  (1927).

\bibitem{Pawellek:2008st}
M.~Pawellek,
``Quantum mass correction for the twisted kink,''
J. Math. Phys. \textbf{42} (2009), 045404
doi:10.1088/1751-8113/42/4/045404
[arXiv:0802.0710 [hep-th]].

\bibitem{dashen:1975xh}
    R.~F.~Dashen, B.~Hasslacher and A.~Neveu,
    ``Semiclassical Bound States in an Asymptotically Free Theory,''
    Phys. Rev. D \textbf{12} (1975), 2443
    doi:10.1103/PhysRevD.12.2443.




\bibitem{Camporesi_1996}
R.~Camporesi and A.~Higuchi,
``On the eigenfunctions of the Dirac operator on spheres and real hyperbolic spaces,''
\textit{J. Geom. Phys.} \textbf{20} (1996) no.1, 1-18
doi:10.1016/0393-0440(95)00042-9

\bibitem{Khare_2004}
A.~Khare,
``Supersymmetry in quantum mechanics,''
\textit{AIP Conf. Proc.} \textbf{744} (2004), 133-165
doi:10.1063/1.1853201

\bibitem{Dunne:1997ia}
G.~V.~Dunne and J.~Feinberg,
``Self isospectral periodic potentials and supersymmetric quantum mechanics,''
\textit{Phys. Rev. D} \textbf{57} (1998), 1271--1276
doi:10.1103/PhysRevD.57.1271
[arXiv:hep-th/9706012].

\bibitem{Kirsten:2003py}
K.~Kirsten and A.~J.~McKane,
``Functional determinants by contour integration methods,''
\textit{Annals Phys.} \textbf{308} (2003), 502--527
doi:10.1016/S0003-4916(03)00149-0
[arXiv:math-ph/0305010].

\bibitem{Dreiner:2008tw}
H.~K.~Dreiner, H.~E.~Haber and S.~P.~Martin,
``Two-component spinor techniques and Feynman rules for quantum field theory and supersymmetry,''
\textit{Phys. Rept.} \textbf{494} (2010), 1-196
doi:10.1016/j.physrep.2010.05.002
[arXiv:0812.1594 [hep-ph]].

\bibitem{Gates:1983nr} S.~J.~Gates, M.~T.~Grisaru, M.~Roček and W.~Siegel, \textit{Superspace, or One Thousand and One Lessons in Supersymmetry}, Frontiers in Physics \textbf{58} (1983) ISBN: 978-0-8053-3161-5 [arXiv:hep-th/0108200].

\bibitem{Ganguly:2006mx}
A.~Ganguly, M.~V.~Ioffe and L.~M.~Nieto,
``A new effective mass Hamiltonian and associated Lamé equation: Bound states,''
\textit{J. Phys. A} \textbf{39} (2006), 14659--14680
doi:10.1088/0305-4470/39/47/010
[arXiv:quant-ph/0610248].

\bibitem{Khare:2001dc}
A.~Khare and U.~Sukhatme,
``Some exact results for midband and zero band gap states of associated Lamé potentials,''
\textit{J. Math. Phys.} \textbf{42} (2001), 5652--5664
doi:10.1063/1.1416487
[arXiv:quant-ph/0105044].

\bibitem{deSparre:1883}
Cte.~de~Sparre,
``Sur l''equation $\frac{d^2 y}{dx^2}+\cdots=0$,''
\textit{Acta Math.} \textbf{3} (1883), 105-140, 289-321
doi:10.1007/BF02422445, 10.1007/BF02422454

\bibitem{Matveev:2006dtv}
V.~B.~Matveev and A.~O.~Smirnov,
``On the link between the Sparre equation and Darboux--Treibich--Verdier equation,''
\textit{Lett. Math. Phys.} \textbf{76} (2006) no.2, 283-295
doi:10.1007/s11005-006-0074-6

\bibitem{Beccaria:2010zn}
M.~Beccaria, G.~V.~Dunne, G.~Macorini, A.~Tirziu and A.~A.~Tseytlin,
``Exact computation of one-loop correction to energy of pulsating strings in $AdS_5 \times S^5$,''
J. Phys. A \textbf{44} (2011), 015404
doi:10.1088/1751-8113/44/1/015404
[arXiv:1009.2318 [hep-th]].

\end{thebibliography}
\end{document}